\documentclass[fleqn,usenatbib]{mnras}

\usepackage{newtxtext,newtxmath}

\usepackage[T1]{fontenc}

\DeclareRobustCommand{\VAN}[3]{#2}
\let\VANthebibliography\thebibliography
\def\thebibliography{\DeclareRobustCommand{\VAN}[3]{##3}\VANthebibliography}

\usepackage{graphicx}	
\usepackage{amsmath}	

\title[SIRIUS V. Formation of ionized bubbles of the ONC]{SIRIUS Project. V. Formation of off-center ionized bubbles associated with Orion Nebula Cluster}

\author[M. S. Fujii et al.]{
Michiko S. Fujii,$^{1}$\thanks{E-mail: fujii@astron.s.u-tokyo.ac.jp (MSF)}
Kohei Hattori$^2$,
Long Wang,$^{3,1,4}$
Yutaka Hirai,$^{5,6,4}$\thanks{JSPS Research Fellow}
Jun Kumamoto$^{1}$,
Yoshito Shimajiri$^2$\newauthor
and Takayuki R. Saitoh$^{7,8}$
\\
$^{1}$Department of Astronomy, Graduate School of Science, The University of Tokyo, 7-3-1 Hongo, Bunkyo-ku, Tokyo 113-0033, Japan\\
$^2$National Astronomical Observatory of Japan, 2-21-1 Osawa, Mitaka, Tokyo, 181-8588, Japan\\
$^{3}$School of Physics and Astronomy, Sun Yat-sen University, Daxue Road, Zhuhai, 519082, China\\
$^{4}$RIKEN Center for Computational Science, 7-1-26 Minatojima-minami-machi, Chuo-ku,
Kobe, Hyogo 650-0047, Japan\\
$^{5}$Department of Physics, University of Notre Dame, 225 Nieuwland Science Hall, Notre Dame, IN 46556, USA\\
$^{6}$Astronomical Institute, Tohoku University, 6-3 Aramaki, Aoba-ku, Sendai, Miyagi 980-8578, Japan\\
$^7$Department of Planetology, Graduate School of Science, Kobe University, 1-1 Rokkodai-cho, Nada-ku, Kobe, Hyogo 657-8501, Japan\\
$^8$Earth-Life Science Institute, Tokyo Institute of Technology, 2-12-1 Ookayama, Meguro-ku, Tokyo 152-8551, Japan
}

\date{Accepted XXX. Received YYY; in original form ZZZ}

\pubyear{2022}

\begin{document}
\label{firstpage}
\pagerange{\pageref{firstpage}--\pageref{lastpage}}
\maketitle

\begin{abstract}
Massive stars born in star clusters terminate star cluster formation by ionizing the surrounding gas. This process is considered to be prevalent in young star clusters containing massive stars. The Orion Nebula is an excellent example associated with a forming star cluster including several massive stars (the Orion Nebula Cluster; ONC) and a 2-pc size H{\sc ii} region (ionized bubble) opening toward the observer; however, the other side is still covered with dense molecular gas. Recent astrometric data acquired by the Gaia satellite revealed the stellar kinematics in this region. By comparing this data with star cluster formation simulation results, we demonstrate that massive stars born in the ONC center were ejected via three-body encounters. Further, orbit analysis indicates that $\theta^2$ Ori A, the second massive star in this region, was ejected from the ONC center toward the observer and is now returning to the cluster center. Such ejected massive stars can form a hole in the dense molecular cloud and contribute to the formation of the 2-pc bubble. Our results demonstrate that the dynamics of massive stars are essential for the formation of star clusters and H{\sc ii} regions that are not always centered by massive stars. 
\end{abstract}




\section{Introduction}

\begin{figure*}
 \begin{center}
  \includegraphics[width=140mm]{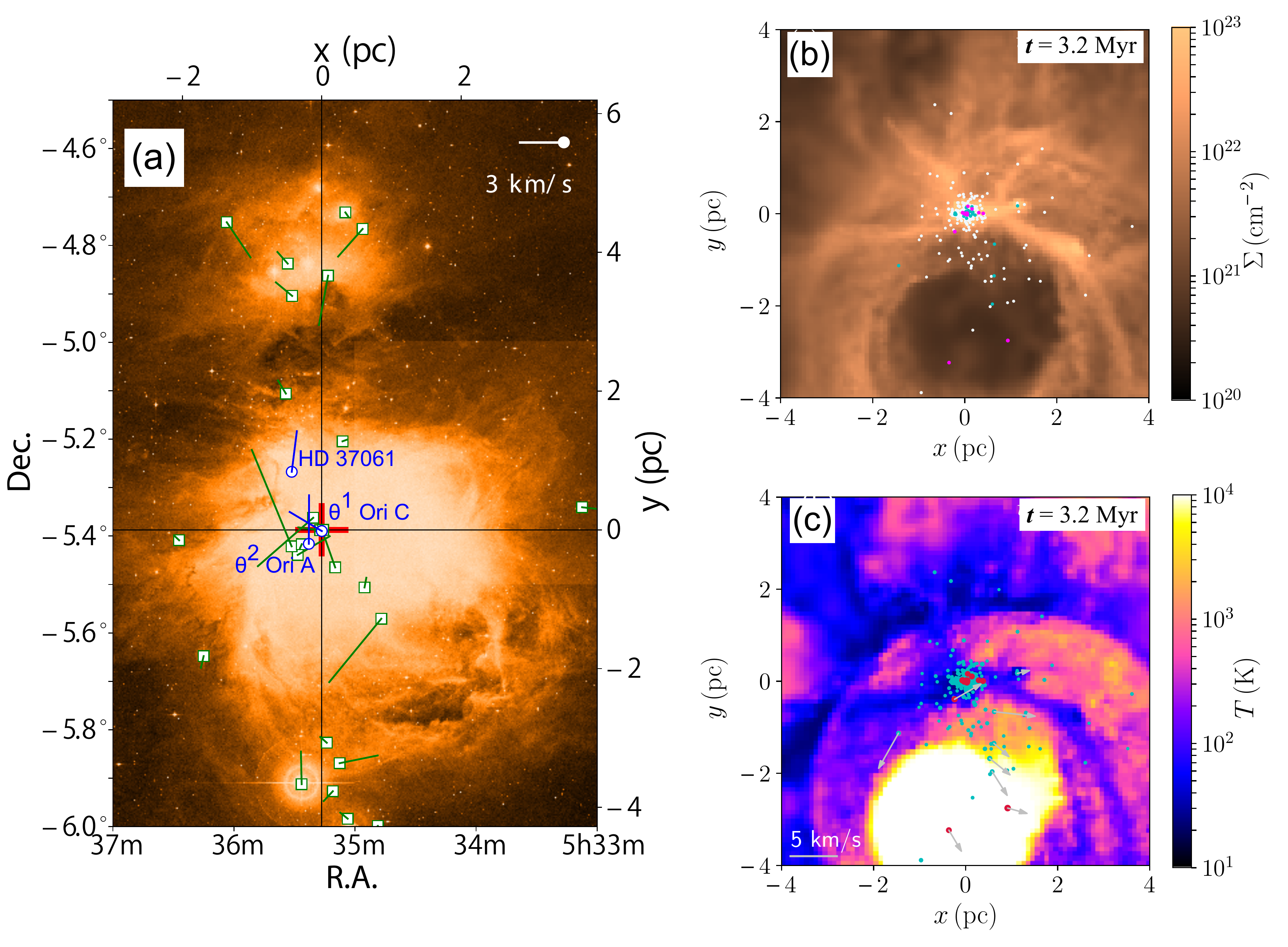}
 \end{center}
\caption{
(a) Distribution of the O-type (blue circles) and B-type (green squares) stars. The three O-stars are NU Ori (HD 37061), $\theta^1$ Ori C, and $\theta^2$ Ori A from top to bottom. The line segment corresponds to the tangential velocity of the stars relative to the ONC (red cross). 
The physical scales $x$ and $y$ in pc are computed assuming that the ONC is 400 pc away from the Sun.
The diffuse component in the background image (red band in the Digitized Sky Survey 2) 
shows the distribution of gas around the ONC region. 
(b) The gas surface density and star distribution in the simulation  (Model A)}. White, cyan, and magenta dots show stars with masses of $>2$, $>10$, and $>20 M_{\odot}$, respectively. (c) The gas temperature of the simulation  Model A between $-1<z<3$\,pc, where the cluster is located at $z=0$. The cyan and red dots show stars with masses of $>2$ and $>20 M_{\odot}$, respectively. The gray arrows indicate the velocities of massive stars.
\label{fig:fig1}
\end{figure*}

The Orion Nebula (M42) is a well-known nebula in the Orion Molecular Cloud and is accompanied by a prominent ionized bubble (H{\sc ii} region, see Fig.~\ref{fig:fig1} (a)). The 2-pc bubble (Extended Orion Nebula) is open toward the observer and considered to be ionized by massive stars in the Orion Nebula Cluster (ONC). The ONC is a young ($\sim 1$\,Myr), massive ($\sim 2000\, M_{\odot}$), and the closest ($388\pm 5$\,pc from us) open cluster located in the brightest region of the Orion Nebula \citep{1998ApJ...492..540H,1999ApJ...525..772P,2017ApJ...834..142K}. The trapezium, a group of massive stars in the dense core of the ONC, illuminates the surrounding gas and forms the H{\sc ii} region \citep{2008Sci...319..309G,2009AJ....137..367O}. In particular, $\theta^1$ Ori C, an O-type star with a mass of 45 solar masses ($M_{\odot}$) \citep{2006A&A...448..351S}, is the most energetic star in the ONC and is, therefore, a candidate source of energy for the formation of the 2-pc bubble \citep{2008Sci...319..309G,2019Natur.565..618P,2021MNRAS.501.1352G}. 

As seen in the ONC, massive stars are closely related to the formation of star clusters. It is considered that the formation of star clusters is terminated by the formation of massive stars within them \citep[][ and references therein]{2019ARA&A..57..227K}. In previous star cluster formation simulations, however, the kinematics of massive stars has not been fully included. The gravitational forces between the stars were softened using a softening length \citep[e.g., ][]{2003MNRAS.343..413B,2021MNRAS.506.5512F}. This treatment prevents close encounters of stars and ejections of stars due to three-body encounters. \citet{2020MNRAS.499..748D} performed simulations of star clusters embedded in a gas cloud including accurate stellar dynamics without gravitational softening. They demonstrated that star clusters are ionized from the outskirts of the clusters due to the dynamical scattering of massive stars in the cluster center. They also reported that this process results in a change in the ionization timescale. Another important influence shown in their simulations is that the ionization bubbles are off-centered when they resolve individual stars and integrate their accurate motions. Thus, the dynamics of massive stars may play an important role in the formation of star clusters.

 Such an off-center bubble is also seen in the Orion Nebula. 
It is considered that the 2-pc ionized bubble associated with the ONC is formed by $\theta^1$ Ori C, which is the most massive star in this region and is located at the center of the ONC. 
The expansion of the ionized bubble is estimated as
\begin{eqnarray}
R_{\rm s}(t)=[125/(154\pi)]^{1/5}(L_{\rm w}/\rho)^{1/5} t^{3/5},\label{eq:shell}
\end{eqnarray}
where $\rho$ is the initial gas density, $L_{\rm w}$ is the mechanical luminosity of the stellar wind, and $t$ is time \citep{1977ApJ...218..377W}. 
Assuming the mechanical luminosity of $\theta^1$ Ori C as $L_{\rm w}=7.5\times 10^{35}$\,erg\,s$^{-1}$ and the density of the molecular cloud as $n\sim10^3$\,cm$^{-3}$, 
\citet{2019Natur.565..618P} estimated that the ionized bubble have expanded to 2.2\,pc in 0.2\,Myr. 
The 2-pc H{\sc ii} region is open toward the observer, and an expanding shell with a speed of $\sim 10$\,km\,s$^{-1}$ is observed \citep{2019Natur.565..618P}.  Thus, $\theta^1$ Ori C is estimated to be energetic enough to form the 2-pc bubble.

Another O-type star observed in the 2-pc bubble is $\theta^2$ Ori A, which is the second massive star located $\sim 0.3$\,pc from the ONC center in the projection \citep{2017ApJ...837..151O}. The mass is estimated to be 25--39\,$M_{\odot}$ \citep{1999NewA....4..531P,2006A&A...448..351S}, and some protoplanetary disks in this region is suggested to be ionized by $\theta^2$ Ori A rather than $\theta^1$ Ori C \citep{2017ApJ...837..151O}. Although its mechanical luminosity is an order of magnitude smaller than that of $\theta^1$ Ori C, $\theta^2$ Ori A could have contributed to the formation of the current structure of the Orion Nebula as seen in the simulation of \citet{2020MNRAS.499..748D}.
 Furthermore, the NU Ori, which is an O-star located at the center of M43 bubble, may also be ejected from the ONC. The dynamical structures of the ONC must be confirmed using the astrometric data in this region.

Recent data releases of Gaia \citep{2021A&A...649A...1G} have revealed the dynamical activities of the ONC. Several runaway ($>30$\,km\,s$^{-1}$) and walkaway ( 5--30\,km\,s$^{-1}$) stars which can be ejected from the ONC has been found \citep{2019ApJ...884....6M,2020MNRAS.495.3104S,2020AJ....159..272P}. 
In our previous paper \citep[][hereafter Paper I]{2021arXiv211115154F}, we performed a simulation of the star cluster formation using our new code \textsc{ASURA+BRIDGE} \citep[SIRIUS project; ][]{2021PASJ...73.1036H,2021PASJ...73.1057F,2021PASJ...73.1074F,2021arXiv211115154F}, with which we can precisely integrate the motion of stars without gravitational softening.
In Paper I, we performed a simulation of a star cluster, which resembles the ONC, and showed its star formation history via hierarchical mergers. We also showed that the numbers of runaway and walkaway stars found around the ONC can be explained using the dynamical ejection of stars in the ONC.

In the present study, we perform a series of simulations for the formation of star clusters starting from a turbulent molecular cloud with \textsc{ASURA+BRIDGE}. We present the dynamical structure of massive stars inside star clusters and compare it with that of the ONC. We also investigate the motions of massive stars in the Orion Nebula using Gaia data.

\section{Methods}

\subsection{Numerical simulation}
We performed the  simulations using a smoothed-particle hydrodynamics code, \textsc{asura+bridge} \citep{2021PASJ...73.1074F}, in which the orbits of stars are integrated with a high-order integrator as collisional systems. \textsc{asura+bridge} combines the hydro part with the stellar part using a Hamiltonian splitting scheme, Bridge \citep{2007PASJ...59.1095F}. In our simulation, all the star particles are integrated using a particle-particle particle-tree (P$^3$T) scheme \citep{2011PASJ...63..881O,2015ComAC...2....6I} using a code for star cluster simulation, \textsc{petar} \citep{2020MNRAS.497..536W,2021PASJ...73.1057F}. In \textsc{petar}, a slow-down algorithmic regularization scheme \citep{2020MNRAS.493.3398W} is included for binaries, and therefore tight binaries can be integrated accurately. \textsc{asura+bridge} includes an HII region model \citep{2021PASJ...73.1074F}, in which the Str\"{o}mgren radius around a massive star is calculated using  the gas distribution around the star. The Str\"{o}mgren radius is iteratively determined to satisfy the balance between the photon count emitted from the massive star and amount of gas absorbing the photons within the radius. The thermal feedback is given to the gas particles within the radius to maintain the gas temperature of $10^4$\,K. Mechanical feedback owing to the stellar wind, which is correlated with the luminosity \citep{2013MNRAS.436.1836R}, is also given to the gas particles within the radius as radial velocities. 

We adopted  initial conditions, which are the same as that used in Paper I (Model A) and smaller cloud model (Model B). Herein, we briefly summarize our initial set-up of the molecular cloud.
We also adopted a homogeneous spherical molecular gas cloud with a turbulent velocity field \citep{2003MNRAS.343..413B}.  For Model A, we set a mass of $2\times 10^4 M_{\odot}$ and a radius of $12$\,pc. The gas particle mass is set as $0.01 M_{\odot}$. Under this condition, the resulting initial gas density and free-fall time are 79.9\,cm$^{-3}$ and 4.87\,Myr, respectively.  For Model B, we adopt a mass of $5\times 10^3 M_{\odot}$ and a radius of 2.285\,pc. The resulting density and free-fall time are $2.89\times 10^3$\,cm$^{-3}$ and 0.81\,Myr, respectively. We adopted $0.1 M_{\odot}$ for the gas particle mass for Model B to reduce the calculation cost. The gas-mass resolution of $0.1 M_{\odot}$ can result in an increase in the formation of stellar mass up to $\sim 10$\,\%, but it does not affect the structures of star clusters. \citep{2021PASJ...73.1074F}.

A turbulent velocity field with power spectrum of $\propto v^{-4}$ was given to the initial condition.  We set the initial virial ratio (kinetic energy / potential energy) as 0.45 and 0.5 for Models A and B, respectively. The turbulent velocity is scaled to satisfy the given virial ratio. These models are similar to that of \citet{2003MNRAS.343..413B}. 
 Because the result depends on the randomness of the initial turbulent velocity field, we performed multiple runs with different random seeds for the turbulence for each model. We performed three and five runs for Models A and B, respectively. We summarize these parameters in Table~\ref{tb:IC}.

For star formation, we used a probabilistic method often used for galaxy simulations; however, we modified it for star formation resolving individual stars \citep{2021PASJ...73.1036H}. In our scheme, stars are formed when gas particles satisfy the following conditions: (1) the gas density exceeds the threshold density ($n_{\rm{th}}$), (2) the gas temperature is lower than the threshold temperature ($T_{\rm{th}}$), and (3) the divergence of the velocity is less than zero. Here we assume $n_{\rm{th}}$ = $7.4\times 10^4$ cm$^{-3}$ and $T_{\rm{th}}$ = 20 K. 
We selected gas particles to star particles following the Schmidt's law \citep{1959ApJ...129..243S}. We thereafter assigned stellar mass following an initial mass function that we assumed. We adopt the Kroupa mass function \citep{2003ApJ...598.1076K}.
As the gas particle mass is always smaller than stellar masses, the new forming stars are created by assembling mass from a given search radius ($r_{\rm{max}}=0.02$\,pc).
Masses of gas particles within $r_{\rm{max}}$ are transported to newly formed stars. 
We limited the stellar mass to be half of the gas mass within $r_{\rm max}$.
If the chosen stellar mass exceeds this limit, we re-assigned the stellar mass until these conditions are satisfied. 

Once star particles are created, we assigned the position and velocity of star particles to ensure momentum conservation. Once the mass of gas particles becomes ten times smaller than its original mass, we merge the gas particle to the nearest gas particles to ensure the number of gas particles is reasonably low. By using this model, we confirmed that it can reproduce the relationship between maximum mass and the enclosed stellar mass \citep{2013MNRAS.434...84W}. This star formation procedure is implemented using a chemical evolution and star formation library, CELib \citep{2017AJ....153...85S}.
These schemes are all tested in previous papers \citep{2021PASJ...73.1036H,2021PASJ...73.1057F,2021PASJ...73.1074F}.

The gravitational forces between gas particles and between gas and star particles are softened with a Plummer type softening, and the softening length ($\epsilon_{\rm g}$) is set to be 0.07\,pc. However, we did not use any softening for star particles. This treatment allows dynamical formation of binaries and scattering of stars owing to three-body encounters. Every 200\,yr (Bridge timestep; $\Delta t_{\rm B}$), stars are perturbed by the gravity of the gas. Moreover, gas particles are integrated using a tree algorithm with hierarchical timesteps. 

We performed the simulation up to 10\,Myr  for Model A and $\sim 4$\,Myr for Model B, at which the molecular cloud is fully ionized.

\begin{table*}
\begin{center}
\caption{{\bf Models and parameters for the simulations of star cluster formation}\label{tb:IC}}
\begin{tabular}{lccccccccccccc}
\hline
   Name  & $M_{\rm g}$  & $m_{\rm g}$ & $R_{\rm g}$& $n_{\rm ini}$ & $t_{\rm ff,ini}$  & $\alpha_{\rm vir}$ & $\epsilon_{\rm g}$ & $\epsilon_{\rm s}$ & $n_{\rm th}$ & $r_{\rm max}$ & $\Delta t_{\rm B}$ & $N_{\rm sin}$ & $N_{\rm run}$\\
       & $(M_{\odot})$ & $(M_{\odot})$ & (pc) &  (cm$^{-3}$) & (Myr) &  & (pc) & (pc) & (cm$^{-3}$) & (pc) & (yr) & &\\ 
      \hline
  Model A & $2\times 10^4$ & $0.01$ & 12 & $79.9$ & $4.87$ & $0.45$  & $0.07$ & $0.0$ & $7.4\times 10^4$ & $0.2$ & $200$ & 1 & 3 \\
  Model B & $5\times 10^3$ & $0.1$ & 2.285 & $2.89\times10^3$ & $0.81$ & $0.5$  & $0.07$ & $0.0$ & $7.4\times 10^4$ & $0.2$ & $200$ & 4 & 5 \\
\hline
\end{tabular}
\end{center}
 From the left: model name, initial cloud mass ($M_{\rm g}$), gas-particle mass ($m_{\rm g}$), initial cloud radius ($R_{\rm g}$), initial cloud density ($n_{\rm ini}$), initial free-fall time ($t_{\rm ff, ini}$), initial virial ratio ($\alpha_{\rm vir}=|E_{\rm k}|/|E_{\rm p}|$), softening length for gas ($\epsilon_{\rm g}$) and stars ($\epsilon_{\rm s}$), star formation threshold density ($n_{\rm th}$), the maximum search radius ($r_{\rm max}$), timestep for Bridge ($\Delta t_{\rm B}$), the number of runs that resulted in the formation of a single star cluster ($N_{\rm sin}$), and the number of runs ($N_{\rm run}$).
\end{table*}

\subsection{OB stars near the ONC from Gaia catalog}
To compare the simulation with the ONC, we compiled a sample of OB stars near the ONC using Gaia Early Data Release 3 (EDR3) in two steps. 
First, we selected all the Gaia sources that satisfy the following properties:
(i) Sky position is within 10$^{\circ}$ from the ONC; 
(ii) Gaia's G-band magnitude is brighter than 12 mag; and
(iii) Gaia's parallax measurement $\varpi \pm \sigma_\varpi$ is consistent with a distance of 340-460 pc within 3 standard deviations. Namely, $\varpi+3\sigma_\varpi < 1/(340 \mathrm{pc})$ and $\varpi-3\sigma_\varpi > 1/(460 \mathrm{pc})$. 
Next, we cross-matched this sample against the SIMBAD catalog 
and selected all the stars that were classified as O- or B-type stars in SIMBAD. 
This procedure resulted in 353 OB stars within 10$^{\circ}$ (corresponding 70\,pc at $d=390$\,pc) from the ONC, 34 of which were within 0.71$^{\circ}$ (5\,pc) from the ONC.

The tangential velocity of OB stars relative to the ONC is given as \\
$(\Delta v_\alpha, \Delta v_\delta)= (4.74047\;\mathrm{km\;s^{-1}}) \times (d / \mathrm{pc}) \times [  (\mu_{\alpha*}, \mu_\delta)-(\mu_{\alpha*}, \mu_\delta)_\mathrm{ONC} ]/(\mathrm{mas\;yr^{-1}})$. 
Here, the heliocentric distance of each star is estimated to be 
$d = 1/\varpi$, where $\varpi$ is the parallax determined by Gaia. 
Most OB stars in our sample had a good parallax measurement; hence, this simple point-estimate of the distance (ignoring the uncertainty) is sufficient for our study. 

Using the proper motion, we calculated the velocities relative to the ONC. The proper motion of the ONC's center-of-mass was assumed to be $(\mu_{\alpha*}, \mu_\delta)_\mathrm{ONC} = (1.1, 0.3)\; \mathrm{mas\;yr^{-1}}$ \citep{2019A&A...627A..57J}. 
We summarized the proper motions and velocity relative to the ONC of the 34 OB stars in Table~\ref{tb:Gaia} in Appendix.

\section{Results}
\subsection{Star cluster formation in the simulation}
\begin{figure*}
 \begin{center}
 \includegraphics[width=70mm]{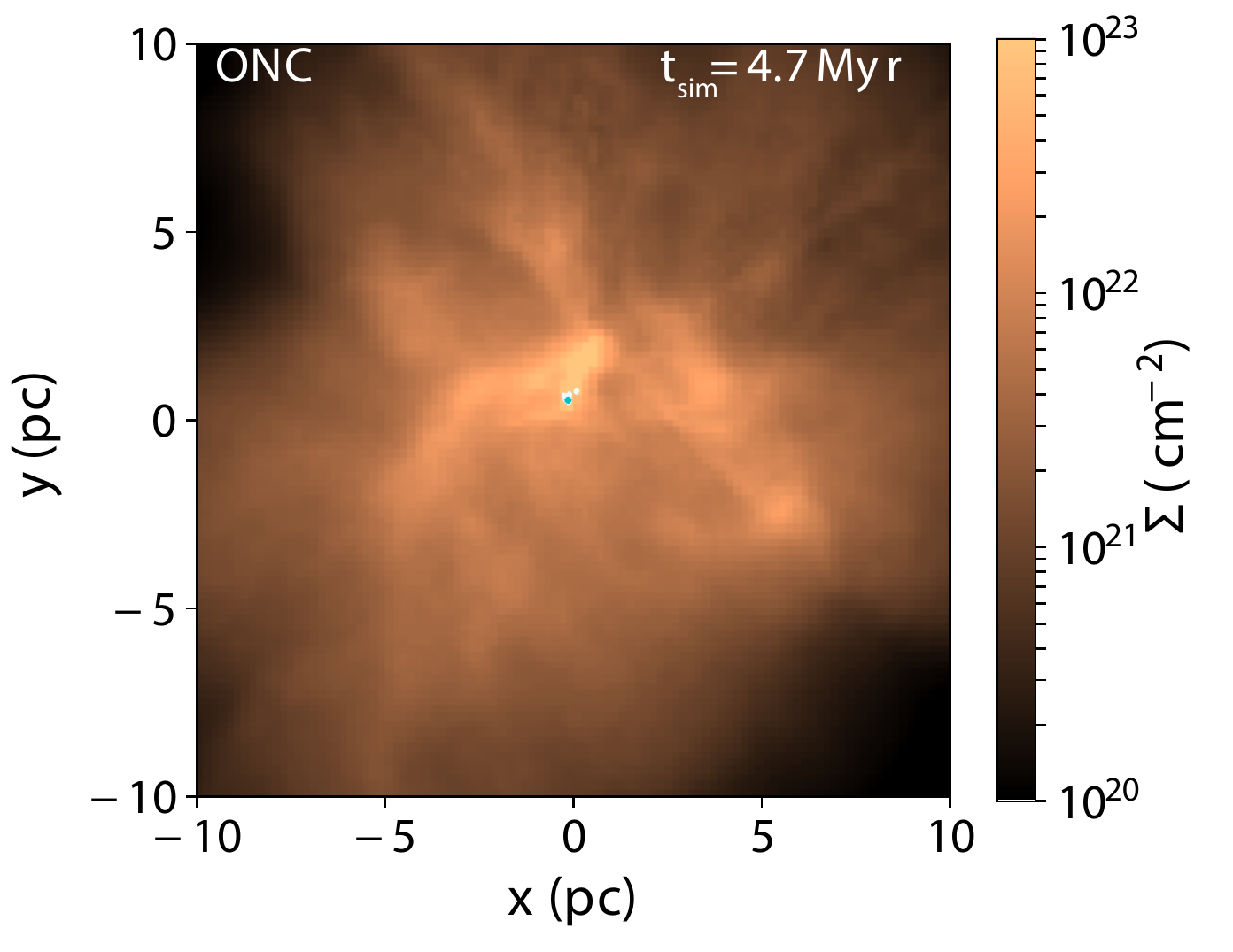}
 \includegraphics[width=70mm]{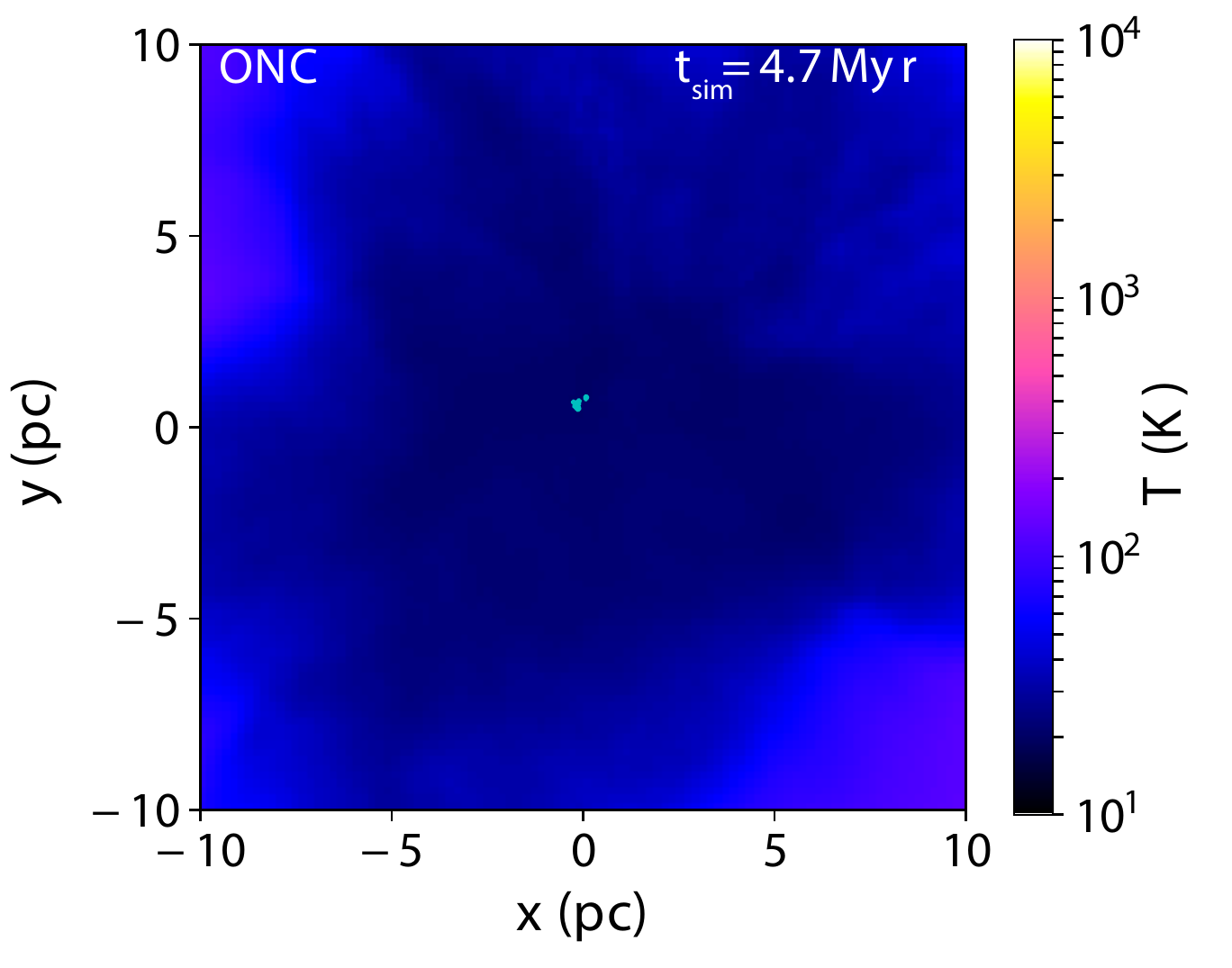}\\
 \includegraphics[width=70mm]{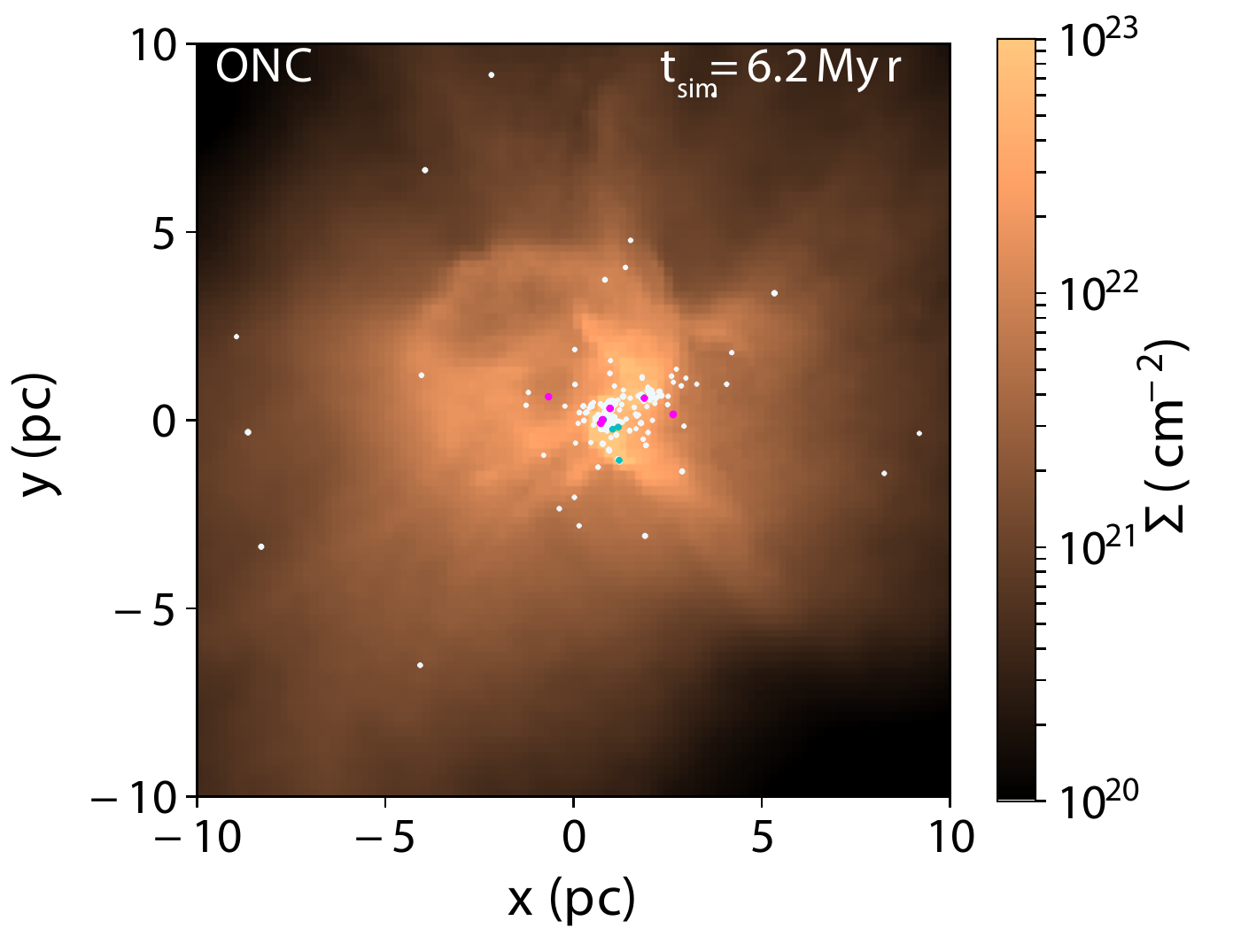}
 \includegraphics[width=70mm]{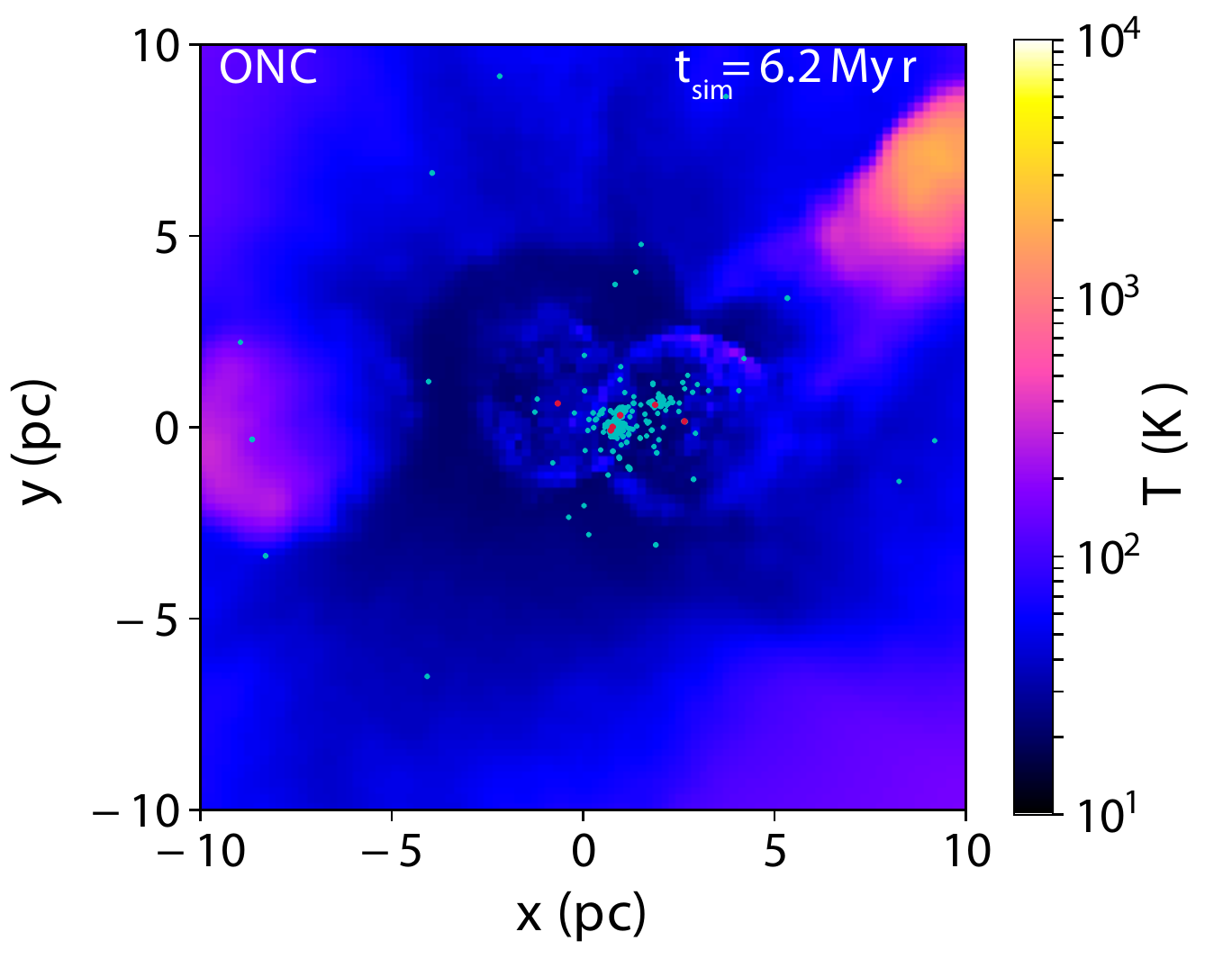}\\
 \includegraphics[width=70mm]{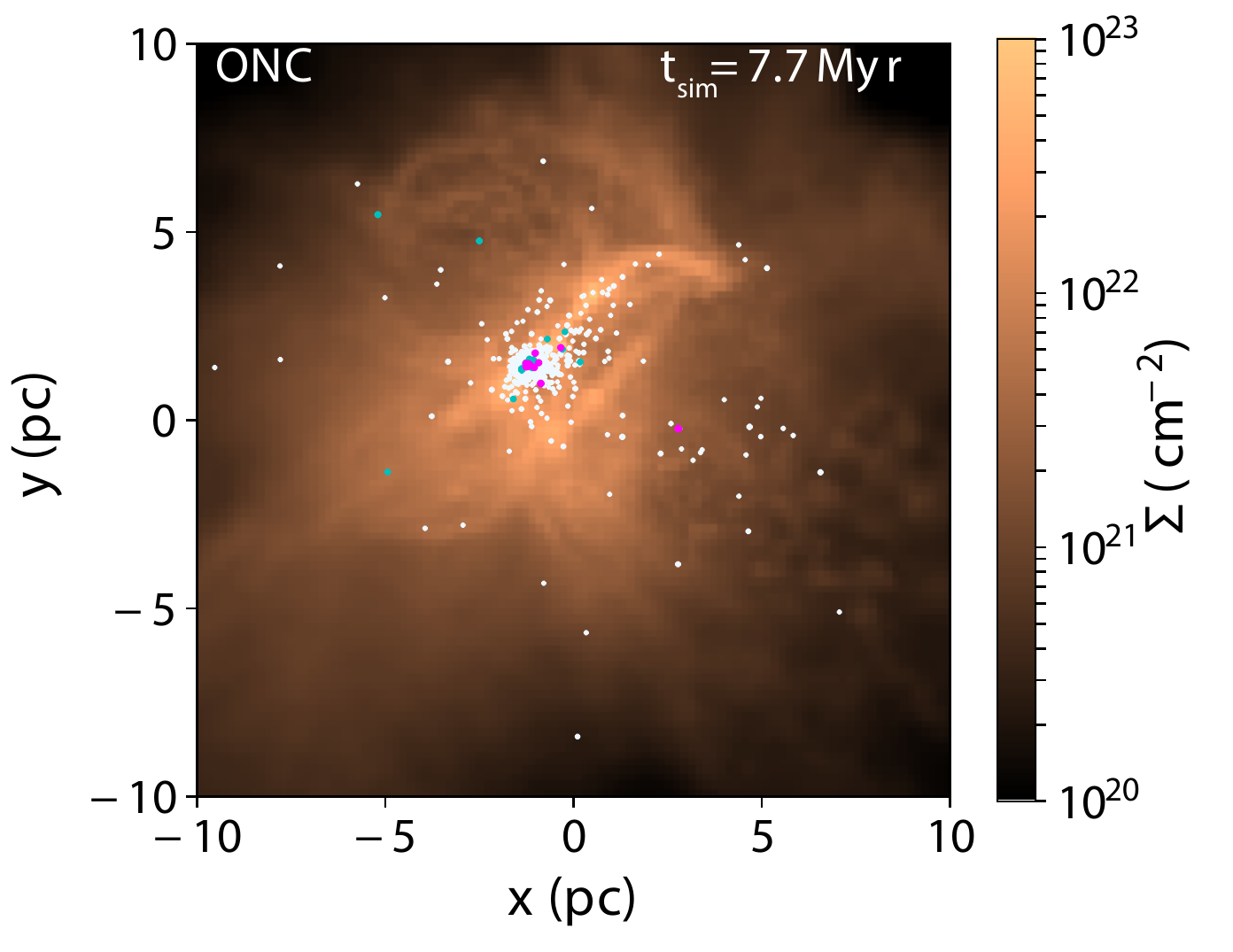}
 \includegraphics[width=70mm]{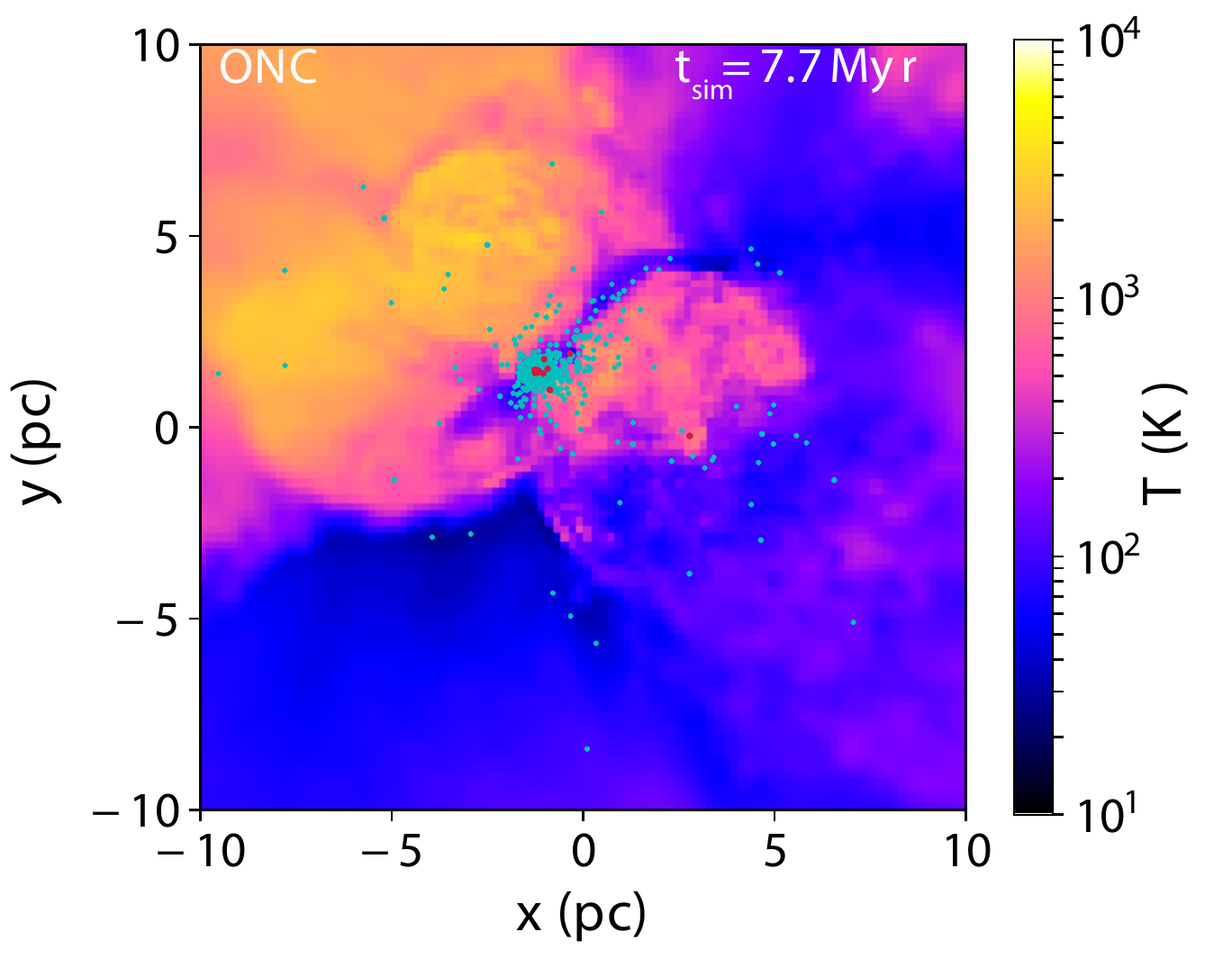}\\
 \includegraphics[width=70mm]{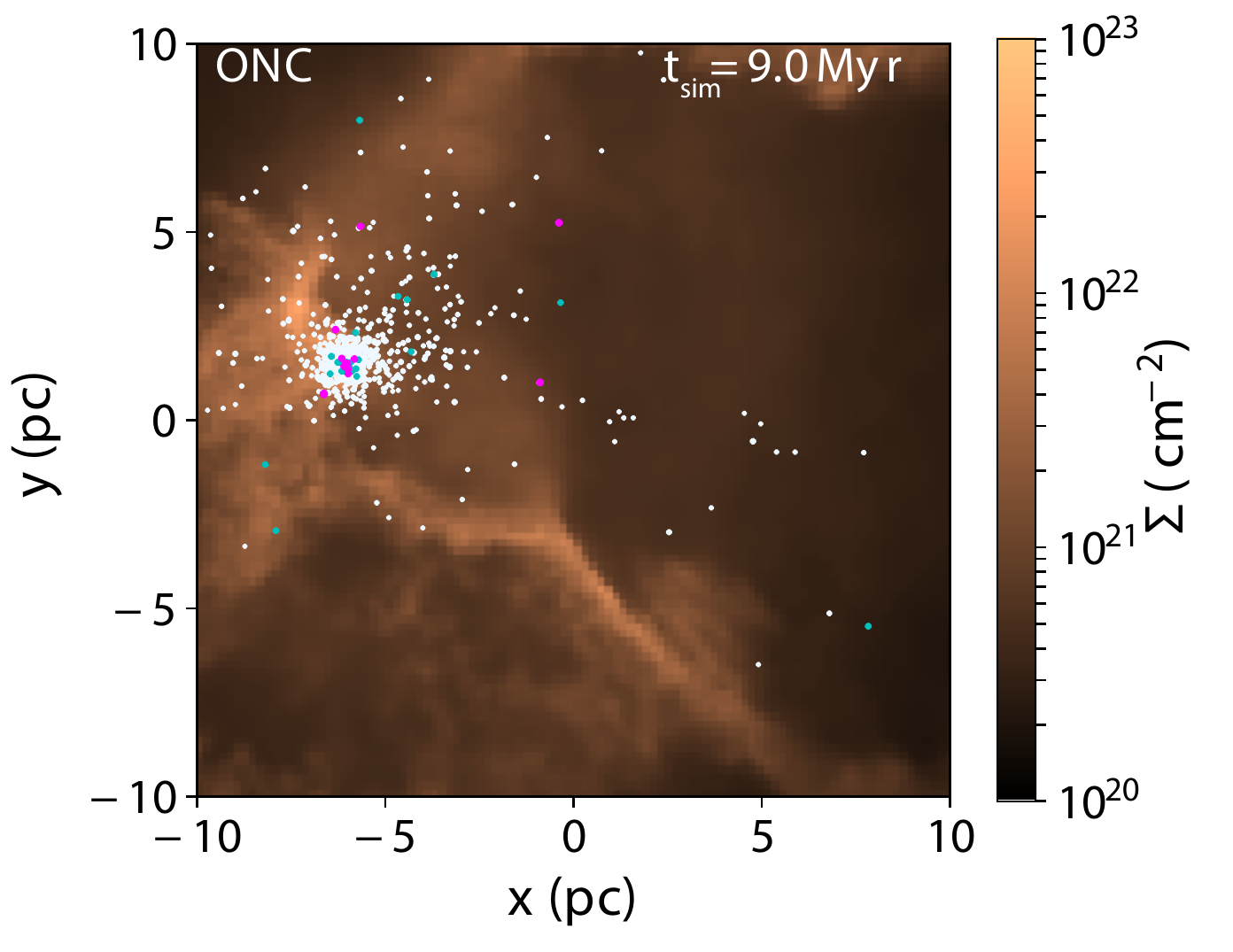}
 \includegraphics[width=70mm]{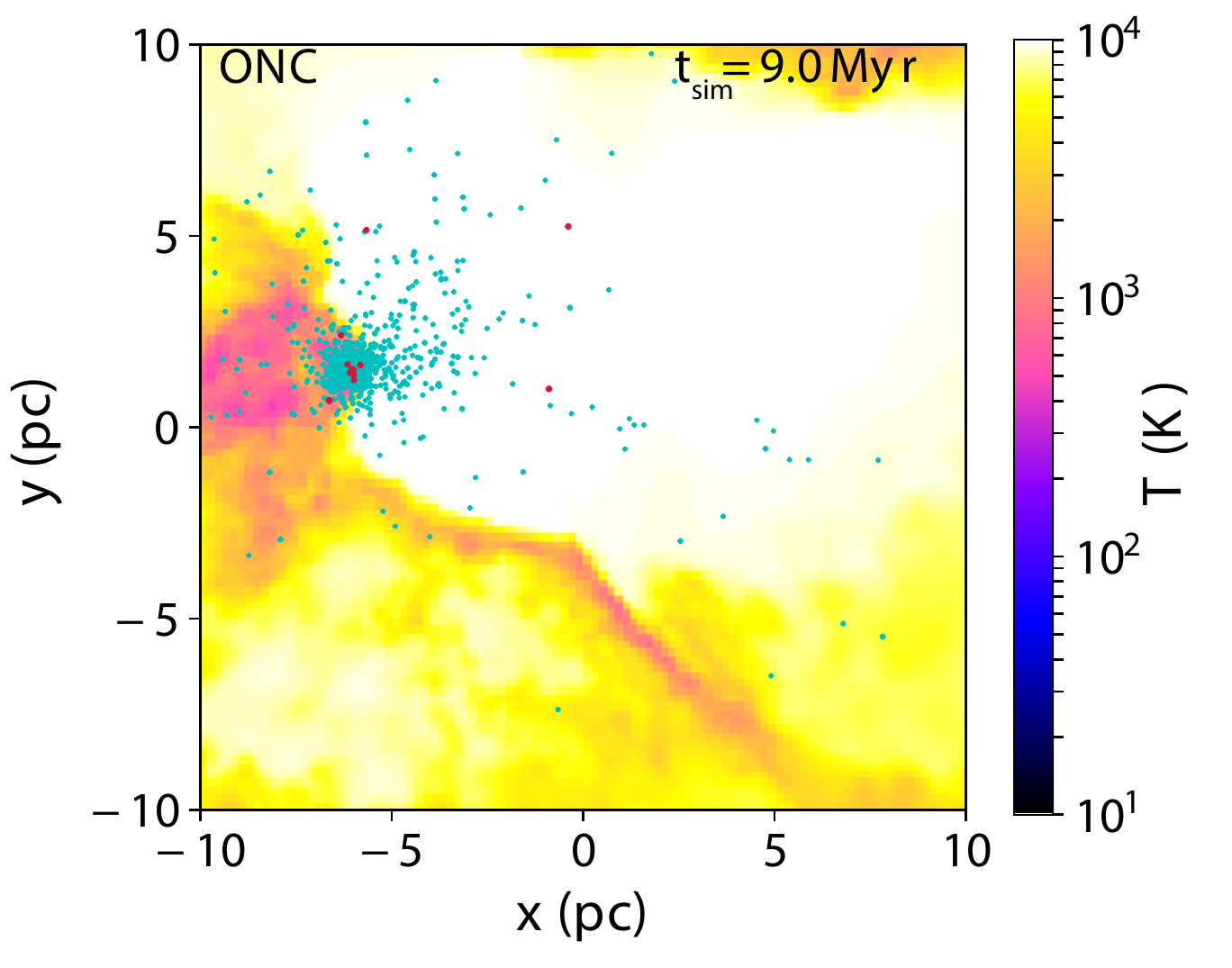}\\
 \end{center}
\caption{Snapshots of the simulation  (Model A)}. Left: gas surface density. Dots indicate stars with $>1 M_{\odot}$. Cyan and red indicate stars with $10<m<20\,M_{\odot}$ and $>20M_{\odot}$, respectively. Right: gas temperature. Cyan dots indicate stars with $>1 M_{\odot}$. Red indicates stars with $>20 M_{\odot
}$. Time indicates the time from the beginning of the simulation.\label{fig:snapshot}
\end{figure*}

In Fig.~\ref{fig:snapshot}, we present the time evolution of gas surface density and temperature distribution with the distribution of stars. The star formation begins  after about one initial free-fall time, which are at 4.87 and 0.81\,Myr for Models A and B, respectively. Hereafter, we indicate the time from the beginning of the simulation as $t_{\rm sim}$ and the time from the first star formation  (4.5 and 0.7 \,Myr for Models A and B, respectively) as $t$. The cluster evolves via mergers of stellar clumps and the outer region is ionized first owing to the dynamical ejection of massive stars. 

\begin{figure}
 \begin{center}
 \includegraphics[width=80mm]{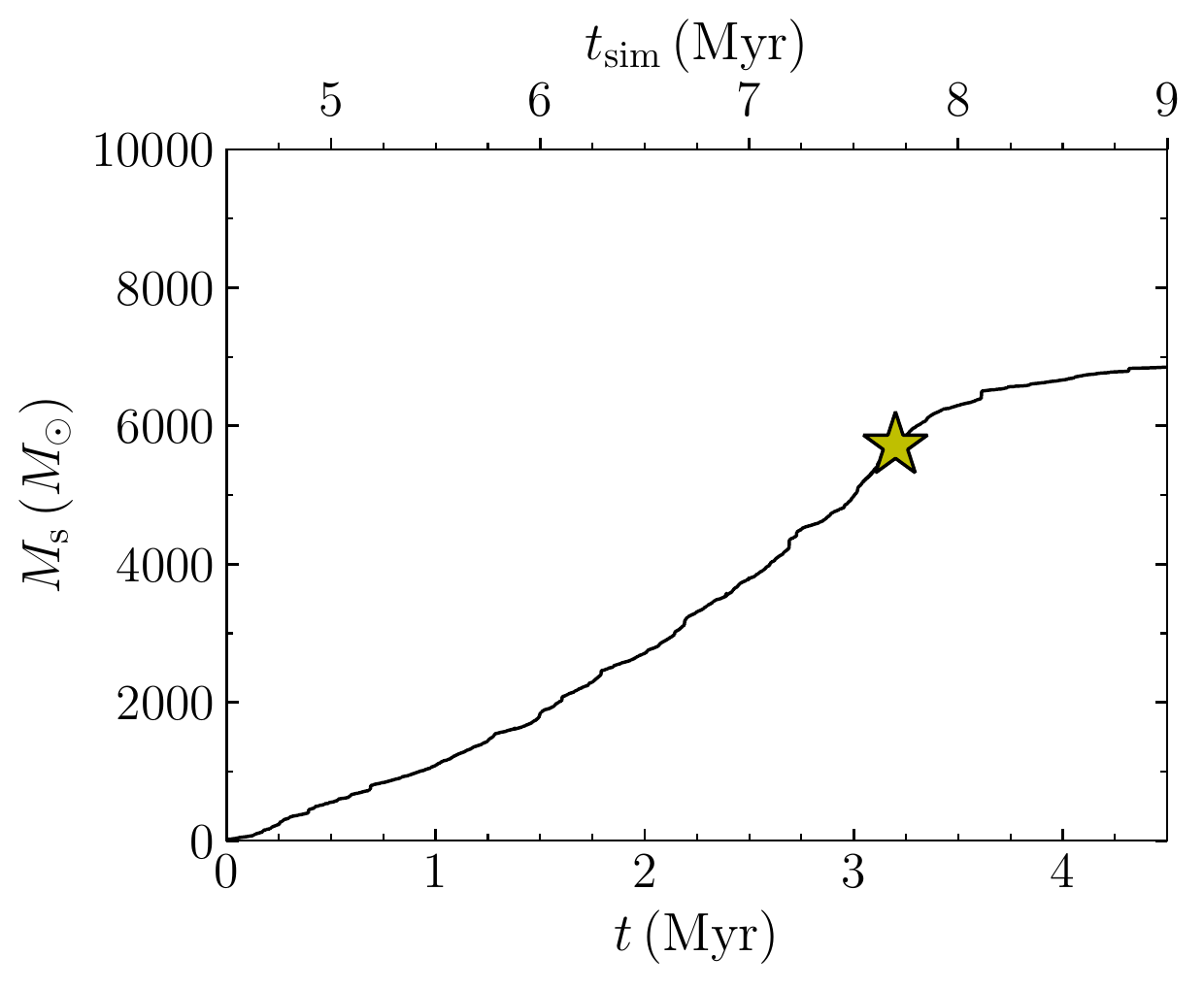}
 \includegraphics[width=80mm]{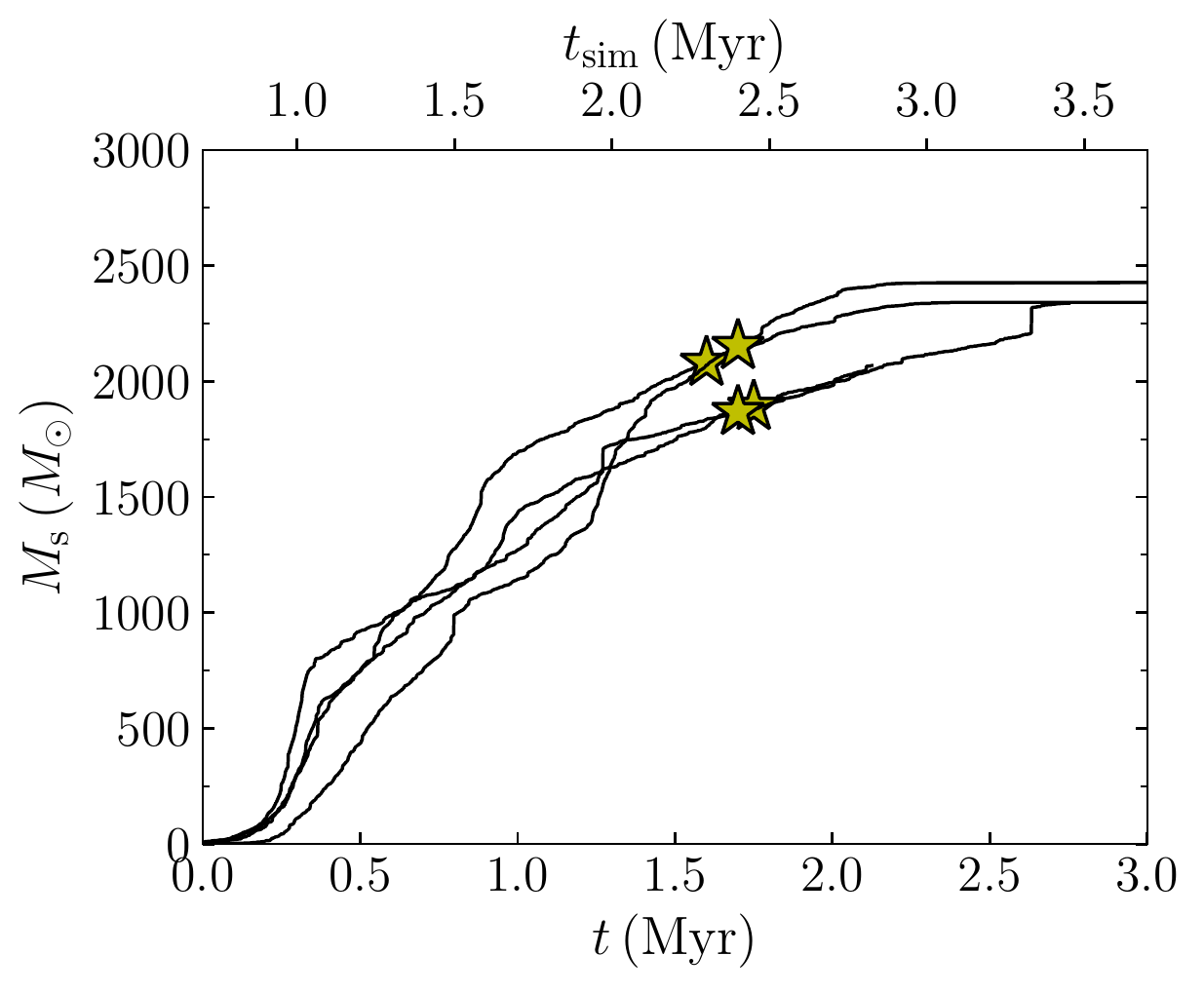}
 \end{center}
\caption{Stellar mass evolution in the simulations ; Model A (top) and Model B (bottom)}. Star symbol indicates the time when we compared to the ONC  (the time at which the gas and stellar masses in the cluster are comparable). 
The $x$-axis shows the time from the first star formation, which is 4.5\,Myr  (Model A) and 0.7 \,Myr (Model B) from the beginning of the simulation (top axis, $t_{\rm sim}$).
\label{fig:stellar_mass_ev}
\end{figure}

The total stellar mass evolution in the simulation is shown in Fig.~\ref{fig:stellar_mass_ev}.  We assumed $t=3.2$\,Myr ($t_{\rm sim}=7.7$\,Myr), at which the stellar and gas masses in the cluster are comparable, as the present day of the ONC for Model . Meanwhile, a part of the gas in the cluster center was ionized; however, some gas remained as dense molecular cloud, and therefore star formation is still ongoing in the central region of the cluster. 
The age of stars in the ONC distributes within 10\,Myr, but the majority were formed within the past 2\,Myr and the star formation rate is accelerated \citep{1999ApJ...525..772P}. Our simulation is consistent with such observational features of the star formation in the ONC. The average stellar  age at this time is 1.7\,Myr, which is also consistent with the age of the ONC (1--3\,Myr) \citep[e.g., ][]{1998ApJ...492..540H,2010ApJ...722.1092D}.

\begin{figure}
 \begin{center}
 \includegraphics[width=80mm]{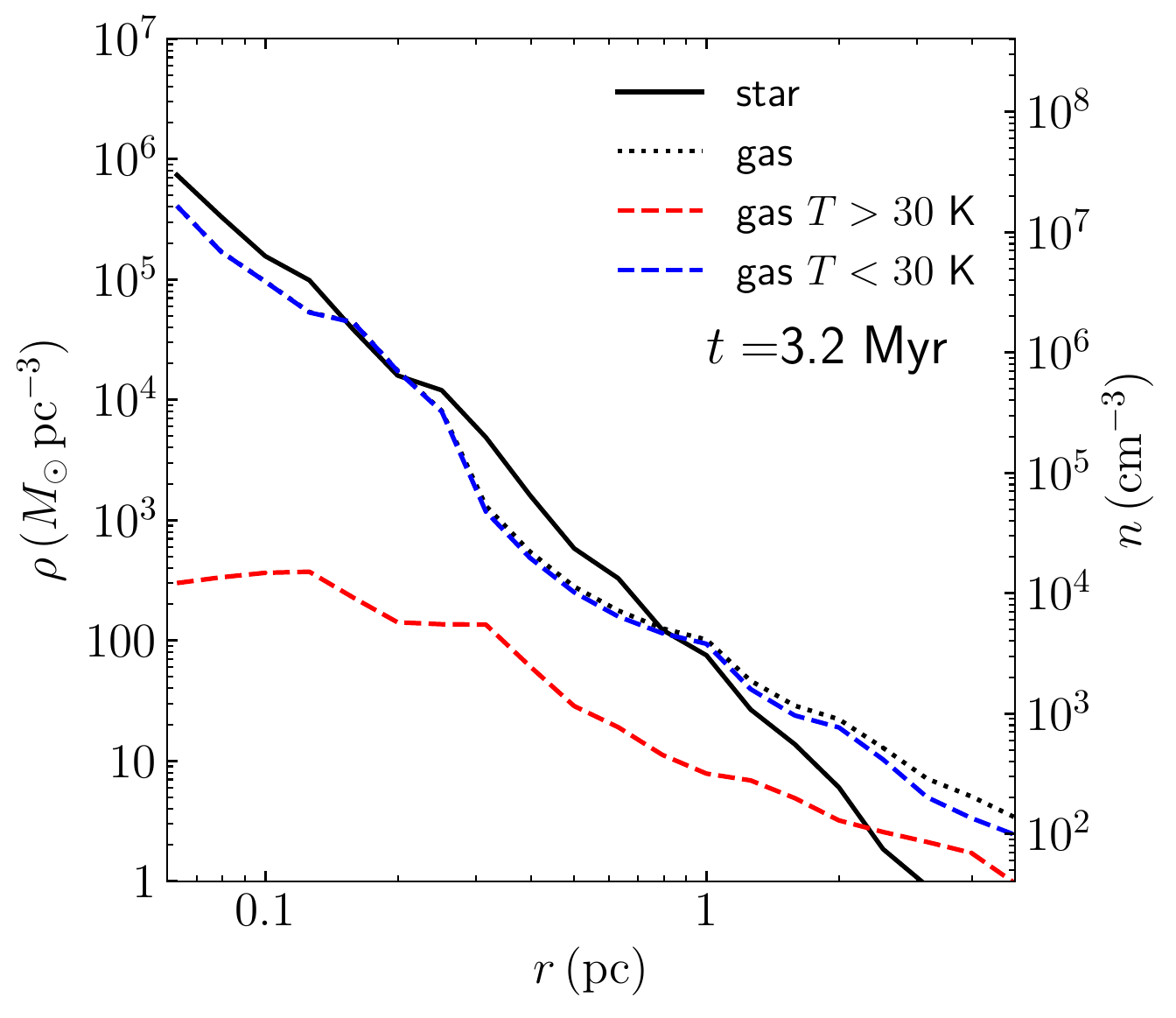}
 \end{center}
\caption{Density profiles of stars and gas at $t=3.2 ~(t_{\rm sim}=7.7)$\,Myr from the beginning of the star formation (simulation)  for Model A}. Black solid and dotted curves indicate the stellar and gas distributions, respectively. Blue and red dashed curves indicate cold gas with a temperature of $<30$\,K and warm and hot gas with a temperature $>30$\,K. \label{fig:densiyt_prof}
\end{figure}

The radial density distributions of gas and stars at $t=3.2$\,Myr are shown in Fig.~\ref{fig:densiyt_prof}.  
The central region within $<0.3$\,pc has a density higher than $10^5$\,cm$^{-3}$, which is sufficient for star formation \citep{2002ApJ...575..950O}. The cold gas density continues to increase at the center of the cluster until $3.2$\,Myr. 
The central density of cold gas reaches $10^7$\,cm$^{-3}$, which is comparable to the observed molecular gas density in the ONC.
The Atacama Large Millimeter/submillimeter Array (ALMA) N$_2$H$^+$ ($J$=1--0) and Atacama Pathfinder EXperiment (APEX) N$_2$H$^{+}$ ($J$=7--6) observations toward ONC revealed that the density of molecular clouds near the ONC center is estimated to reach $10^6$\,cm$^{-3}$ and can partially exceed $10^7$\,cm$^{-3}$ \citep{2020A&A...644A.133H}. 

After $t\sim 3.5$\,Myr, the star formation slows down (see Fig.~\ref{fig:stellar_mass_ev}), and at a later stage, the central cold gas density decreases. At the end of this simulation ($t=$4.5\,Myr), the amount of cold gas in the cluster center is only $\sim 10 M_{\odot}$, and it can no longer form new stars.

Once massive stars form in the simulation, the feedback ionizes and blows away the surrounding gas (see Fig.~\ref{fig:snapshot}). However, it takes a few Myr for the massive stars to fully ionize the molecular cloud. 
In the panels (b) and (c) of Fig.~\ref{fig:fig1}, the snapshots at $t=3.2$\,Myr are shown. The total stellar mass of the simulated cluster at this time was $5200 M_{\odot}$ within 3\,pc from the cluster center (see Table~\ref{tb:result}).  
In these snapshots, an ionized bubble similar to the Orion Nebula is observed, and the star formation is still ongoing in the dense molecular cloud close to the cluster center. These structures are similar to those of the Orion Nebula (see Fig.~\ref{fig:fig1} (a)). 
This bubble is formed by a star ejected from the cluster center due to strong dynamical interactions such as three-body encounters. 

 The mass of the star clusters that are formed and the number of star clusters change depending on the randomness of the initial turbulent velocity field. Although we performed three and five runs for Models A and B, respectively, one and four of them resulted in the formation of single star clusters, which is similar to the ONC. The other runs resulted in twin or multiple star clusters. We exclude these runs from the following analyses. 

\begin{table}
\begin{center}
\caption{ Simulated star clusters}\label{tb:result}
\begin{tabular}{lccccc}
\hline
   Name  & $t_{\rm ONC, sim}$ & $t_{\rm ONC}$ & $M_{\rm s, 3pc}$  & $M_{\rm g, 3pc}$ & $v_{\rm esc}$\\
       & (Myr) & (Myr) & $(M_{\odot})$ & $(M_{\odot})$ & (km\,s$^{-1}$) \\ 
      \hline
  Model A & $7.7$ & $3.2$ & 5200 & 4540 & 5.29  \\
  Model B1 & 2.3 & 1.6 & 1590 & 1580 & 3.01 \\
  Model B2 & 2.45 & 1.75 & 1630 & 1200 & 2.85 \\
  Model B3 & 2.4 & 1.7 & 2000 & 1990 & 3.38 \\
  Model B4 & 2.4 & 1.7 & 1630 & 1640 & 3.07 \\
\hline
\end{tabular}
\end{center}
 From the left: model name, time at which we compare the model with the ONC from the beginning of the simulation and the star formation ($t_{\rm ONC, sim}$ and $t_{\rm ONC}$), stellar and gas masses within 3\,pc from the cluster center ($M_{\rm s, 3pc}$ and $M_{\rm g, 3pc}$), and escape velocity at 3\,pc ($v_{\rm esc}$). The numbers in the model names indicate the runs with different random seed for the initial turbulent velocity field.
\end{table}

\subsection{Velocity distribution of stars around the ONC}

If the ONC has been formed via star formation in turbulent molecular clouds, stars formed in the central region of the cluster must be scattered from the cluster center, and this mechanism must be confirmed by the velocity distribution of stars around the ONC. 
Therefore, we compared the velocity distribution of the stars within 5\,pc from the ONC to that obtained from the simulation. The Gaia astrometry satellite has added kinematic data (positions and velocities) of stars in this region (see Fig.~\ref{fig:fig1} (a)), which provide insights into the dynamical evolution of this region \citep{2021A&A...649A...1G}. 

\begin{figure}
 \begin{center}
 \includegraphics[width=80mm]{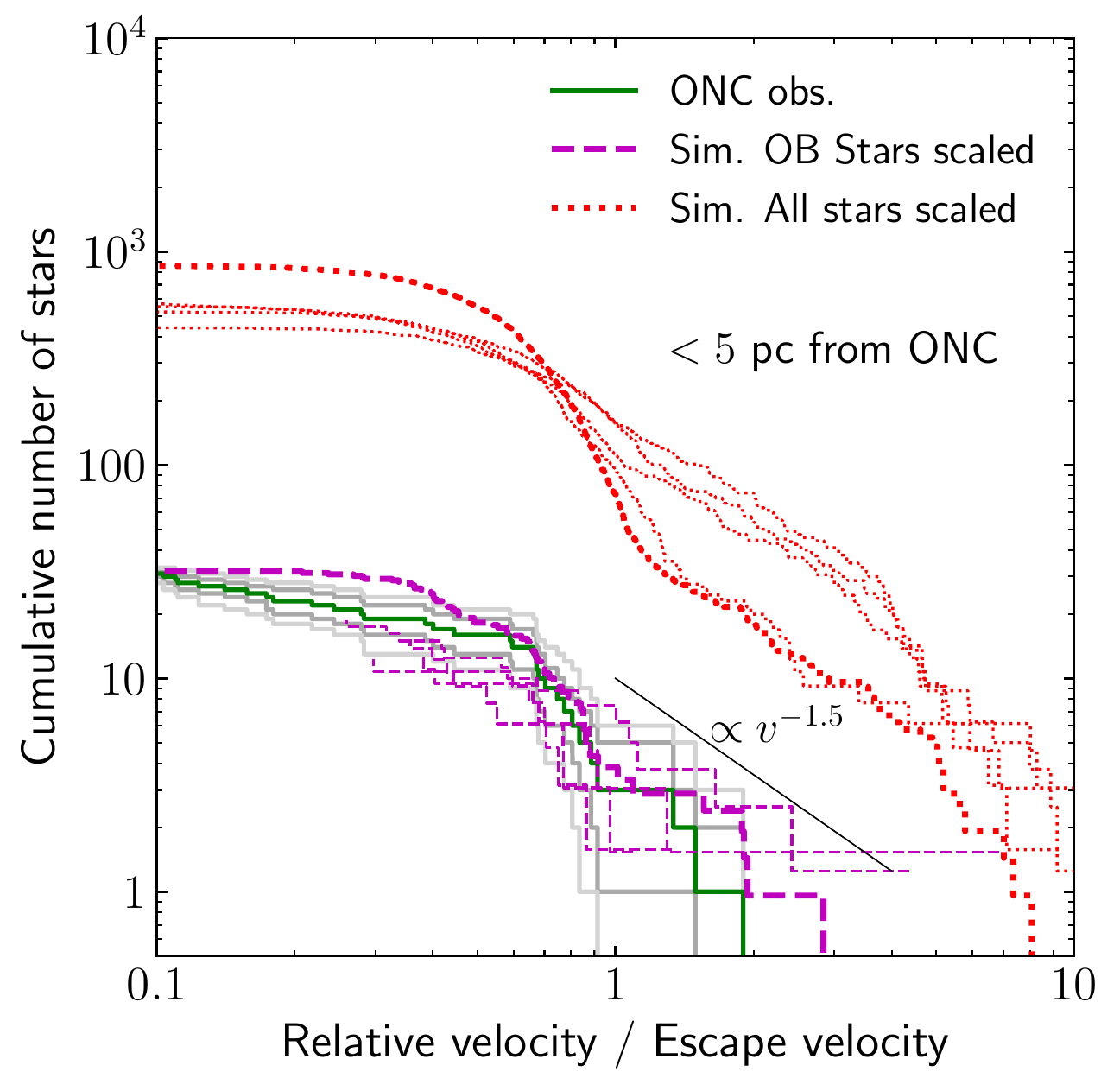}
 \end{center}
\caption{Velocity distribution of stars within 5 pc from the  star cluster}. The green curve represents the OB stars observed near the ONC (50-th percentile). The gray curves indicate
the 2.5-, 16-, 84-, and 97.5-percentiles (see the text). The red dotted and magenta dashed curves indicate stars observed in our simulations between 0.8 and 5 pc from the center of the formed cluster. Here, the velocities in the simulation are multiplied by $\sqrt{2/3}$.  The velocities of simulations are scaled by the escape velocities measured for both the observation and simulation. We used the total (gas + stellar) masses within 3\,pc of $5000\,M_{\odot}$ \citep{1998ApJ...492..540H,2006ApJ...641L.121T} for the observation.  The cumulative numbers of stars are also scaled to the ONC mass ($2500 M_{\odot}$) for the simulations. The total mass and escape velocities of the simulations are summarized in Table~\ref{tb:result}. Thick and thin curves indicate Model A and B, respectively. 
\label{fig:vel_dist}
\end{figure}

Fig.~\ref{fig:vel_dist} shows the cumulative velocity distribution of the massive stars ($>2 M_{\odot}$) within 5\,pc from the clusters for the simulations and observation. 
 We compare our simulations to the observations at the time when stellar and gas mass inside 3\,pc from the cluster center are comparable.
The velocities of the stars with respect to the ONC are obtained from Gaia data \citep{2021A&A...649A...1G}.
The total stellar mass of the ONC is estimated to be $1800$--$2700 M_{\odot}$, and the total mass including the gas within $\sim 3$\,pc is estimated to be $\sim 5000 M_{\odot}$ \citep{1998ApJ...492..540H,2006ApJ...641L.121T}. 
 Because the total masses of the clusters in the simulations are not exactly the same as that of the ONC, we scaled the number of stars and velocity relative to the cluster by the total mass and escape velocity of the cluster, respectively. The cluster and gas masses at the time we adopted are summarized in Table~\ref{tb:result}. While Model A was twice as big as the ONC, all of Model B were less massive.

As the number of observed OB stars is only 34, we evaluate the Poisson noise in the cumulative tangential velocity distribution of the 34 OB stars near the ONC (see Table~\ref{tb:Gaia}). 
First, we denoted the magnitude of the tangential velocity as $v_{\mathrm{obs},n}$ for $n$th star in our sample ($n=1,\cdots,34$), without taking into account the observational error. 
Then, we created 100000 bootstrapped samples 
$\{v^{(j)}_i \mid i= 1,\cdots,34 \}$ where $j=1,\cdots,100000$ denotes $j$th bootstrap sample. 
We note that $i$th star in $j$th bootstrap sample is randomly chosen from $\{ v_{\mathrm{obs},n} \}$, and that we are allowed to choose the same star multiple times. 
For $j$th bootstrap sample, the cumulative distribution of $\{v^{(j)}_i \}$ is expressed as $CDF^{(j)}(v)$. 
For any given value of velocity $v$, we computed the 2.5-, 16-, 50-, 84-, and 97.5-percentile points of $\{CDF^{(j)}(v) \mid  j=1,\cdots,100000\}$ using the 100000 cumulative distribution functions. 
By varying $v$, we evaluated how these percentile points change as a function of $v$.  We present the result in Fig.~\ref{fig:vel_dist}.

The velocity distributions of the massive stars were identical for both the simulations and observation.
The number of stars drops around the escape velocity; however, the distribution continues beyond the escape velocity following a power law of $-1.5$. This power-law distribution is formed with high-velocity stars scattered inside the star cluster owing to three-body encounters \citep{2012ApJ...751..133P}. Thus, the velocity distribution of stars around the ONC suggests that scattering of massive stars occurred in the ONC. 

Some stars were ejected from the cluster with a velocity higher than 30 km\,s$^{-1}$. They escape from the cluster and can be recognized as runaway stars. Slower stars (5--30\,km\,s$^{-1}$) are called as walkaway stars. 
Notably, some runaway and walkaway candidates originated from the ONC have been found \citep{2019ApJ...884....6M,2020MNRAS.495.3104S,2020AJ....159..272P}. We quantitatively discussed the runaway and walkaway stars around the ONC in Paper I.

\subsection{Ionization due to scattered massive stars}

\begin{figure*}
 \begin{center}
  \includegraphics[width=180mm]{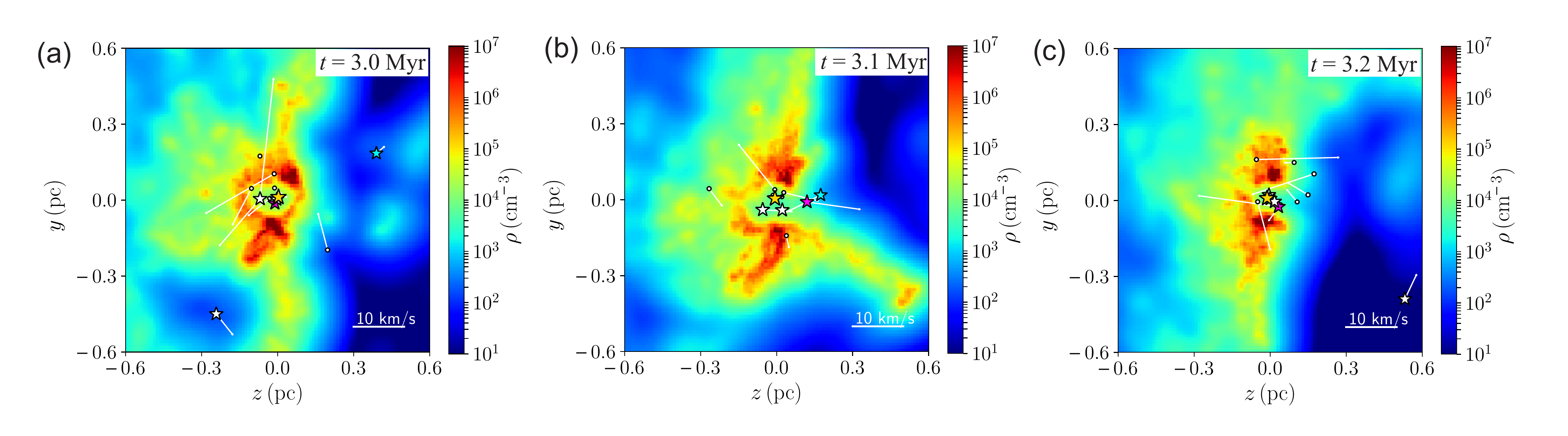}
 \end{center}
\caption{Edge-on density distribution of the molecular gas ($T<100$\,K) at the central region of the cluster in  Model A} for $t=3.0$, 3.1, and 3.2\,Myr from left to right. Panel (c) is the time same for the snapshots shown in Fig.~\ref{fig:fig1}. The $z$-axis is the direction of the observer. The star symbols are massive stars with $>16 M_{\odot}$ within $x=\pm 0.3$\,pc. Yellow star is the most massive star in this region. Magenta and cyan stars are stars which are scattered from the cluster center and then return.\label{fig:CO_dist}
\end{figure*}

The scattered massive stars contribute to the formation of the off-center ionized bubbles. In Fig.~\ref{fig:CO_dist}, we show the distributions of massive stars and cold ($<100$\,K) gas. The right side of this panel is the $z$-direction in the right panels of Fig.~\ref{fig:fig1}. As shown in this figure, star cluster is formed in the center of filament along the $y$-axis, and in the innermost region, massive stars form a $0.1$\,pc-scale H{\sc ii} region. 

Initially (panel (a)), the inner 0.1\,pc-bubble is covered with molecular gas. However, in panels (b) and (c), the small bubble is open toward the observer. This can be attributed to the ejection of a massive star from the central region. 
This structure is similar to that observed in the Orion Nebula \citep{2009AJ....137..367O,2019ApJ...881..130A,2020ApJ...891...46O}. Behind the 2\,pc bubble, small 0.1\,pc-scale bubble exists.

Another O-type star observed in the 2-pc bubble is $\theta^2$ Ori A, which is the second massive star located $\sim 0.3$\,pc from the ONC center in the projection \citep{2017ApJ...837..151O}. The mass is estimated to be 25--39\,$M_{\odot}$ \citep{1999NewA....4..531P,2006A&A...448..351S}, and some protoplanetary discs in this region is suggested to be ionized by $\theta^2$ Ori A rather than $\theta^1$ Ori C \citep{2017ApJ...837..151O}. We estimated the mechanical luminosity of $\theta^2$ Ori A to be $7.4\times 10^{34}$\,erg\,s$^{-1}$ (see Table~\ref{tb:stars} and Appendix A). Although it is an order of magnitude smaller than that of $\theta^1$ Ori C, the bubble expansion time to 0.1\,pc driven by $\theta^2$ Ori A is only 0.05\,Myr for $n=10^7$\,cm$^{-3}$. This was obtained using equation (\ref{eq:shell}). This expansion time is short enough for $\theta^2$ Ori A to partially ionize the filament's wall when it travels to the outer region due to the dynamical ejection.

We estimated the trajectory of $\theta^2$ Ori A using its astrometric data. 
According to Gaia data, this star is located at a distance of $336^{+26}_{-22}$\,pc ($1\sigma$ error) from the observer. This distance seems to be much closer to the observer than the 2-pc bubble, but we note that the parallax error of this star is $\sim 10$\,\%, which is an order of magnitude larger than the other stars (see Table \ref{tb:Gaia}). Owing to $2\sigma$ error, this star is located at a distance of $295$--$391$\,pc. The velocity relative to the ONC is $3.3$\,km\,s$^{-1}$ toward the ONC (see Fig.~\ref{fig:fig1}). The proper motion suggests that $\theta^2$ Ori A was ejected from the ONC center and is currently returning to the ONC. $\theta^2$ Ori A could break the molecular cloud toward the observer. 

Assuming the mass distribution of the ONC, we calculate possible orbits of $\theta^2$ Ori A. 
By fitting a double power-law function to the mass distribution in Fig.~\ref{fig:densiyt_prof} and scaling the mass to the ONC mass, we obtained a mass distribution of $M(<r)=25000\,(r/{\rm 1 pc})^{1.3} M_{\odot}$ for $r<0.175\,{\rm pc}$ and $M(<r)=4000\,(r/{\rm 1pc})^{0.25} M_{\odot}$ for $r>0.175\,{\rm pc}$. 
The enclosed mass at 3\,pc was $5300 M_{\odot}$, which is consistent with the observed ONC mass including gas \citep{1998ApJ...492..540H,2006ApJ...641L.121T}.
We set the initial position as 0.01\,pc and integrated the radial motions of stars by changing the initial radial velocity from 21.4 to 22.2 km\,s$^{-1}$. In this velocity range, the orbit satisfied 0.3\,pc and $-3$\,km\,s$^{-1}$ in projection within 0.1--0.4\,Myr.

\begin{figure}
 \begin{center}
  \includegraphics[width=80mm]{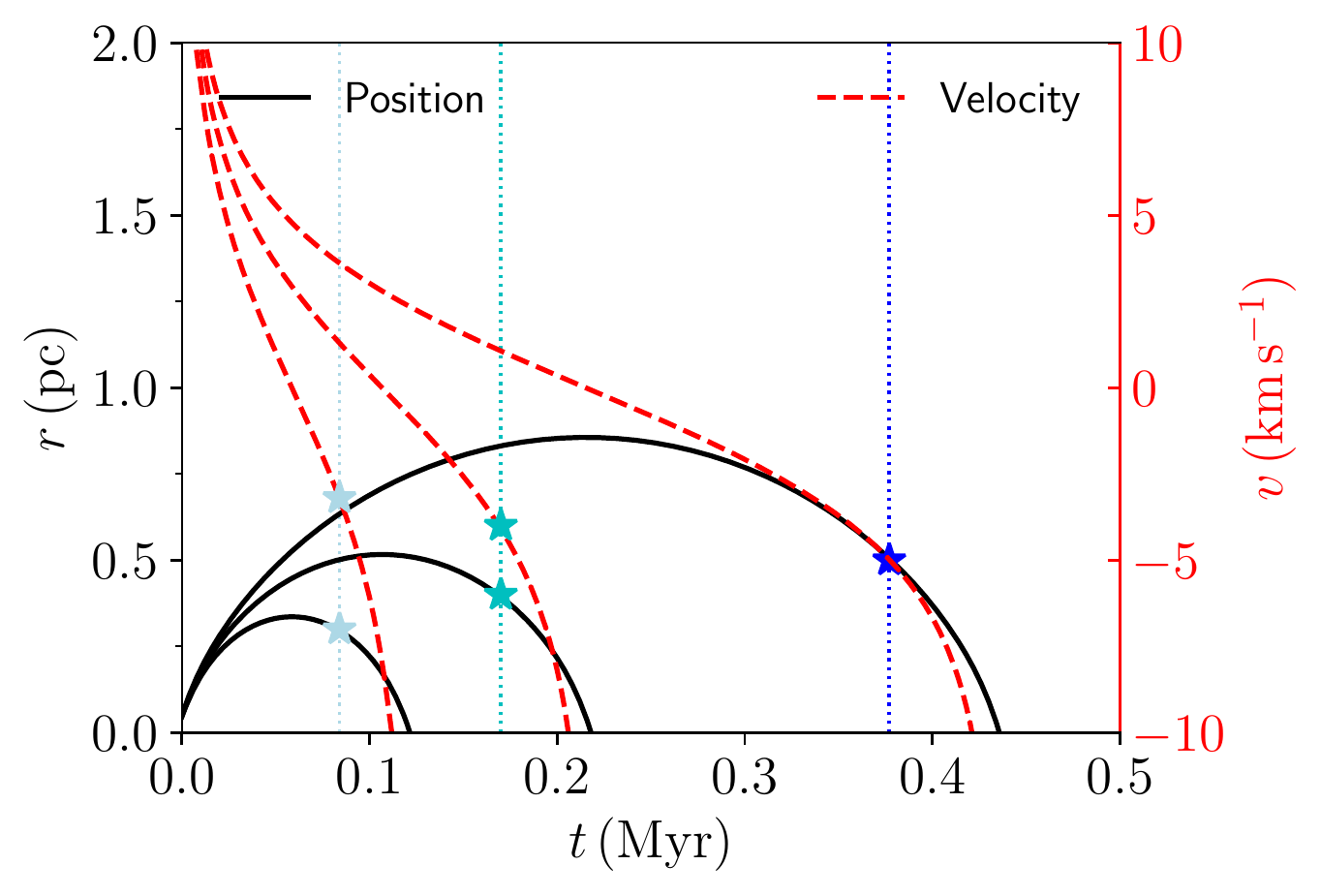}
 \end{center}
\caption{
Possible orbital evolution of stars ejected from the ONC center. Black and red curves indicate the distances from the ONC center and the velocities relative to the ONC, respectively. The curves are for different initial velocities. Blue vertical lines indicate times (0.08, 0.17, and 0.38 Myr) at which the current projected distance (0.3\,pc) and the current projected velocity ($-3$\,km\,s$^{-1}$) of the $\theta^2$ Ori A are satisfied. Stars indicate the positions and velocities at these times.
}\label{fig:radial_motion}
\end{figure}

Some possible orbits (radial distances and velocities) are shown in Fig.~\ref{fig:radial_motion}.
With the current distance of the $\theta^2$ Ori A from the ONC center of $\sim 0.3$\,pc and the velocity to the ONC of 3\,km\,s$^{-1}$, we obtained an ejection time shorter than $\sim 0.4$\,Myr. This timescale is consistent with the age of the 2-pc bubble, which was estimated from the shell velocity ($\sim 0.2$\,Myr) \citep{2019Natur.565..618P,2020A&A...639A...2P}. 

Once $\theta^2$ Ori A was ejected from the ONC center, it contributes to the ionization of the off-center region of the ONC. Assuming the gas density of the outskirt of the cluster as $10^{3}$--$10^4$\,cm$^{-3}$ (see Fig.~\ref{fig:densiyt_prof}), we obtained the bubble size of 1--2\,pc for 0.3\,Myr. In \citet{2017ApJ...837..151O}, some protoplanetary discs are suggested to be ionized by $\theta^2$ Ori A rather than $\theta^1$ Ori C.

\begin{figure}
 \begin{center}
  \includegraphics[width=80mm]{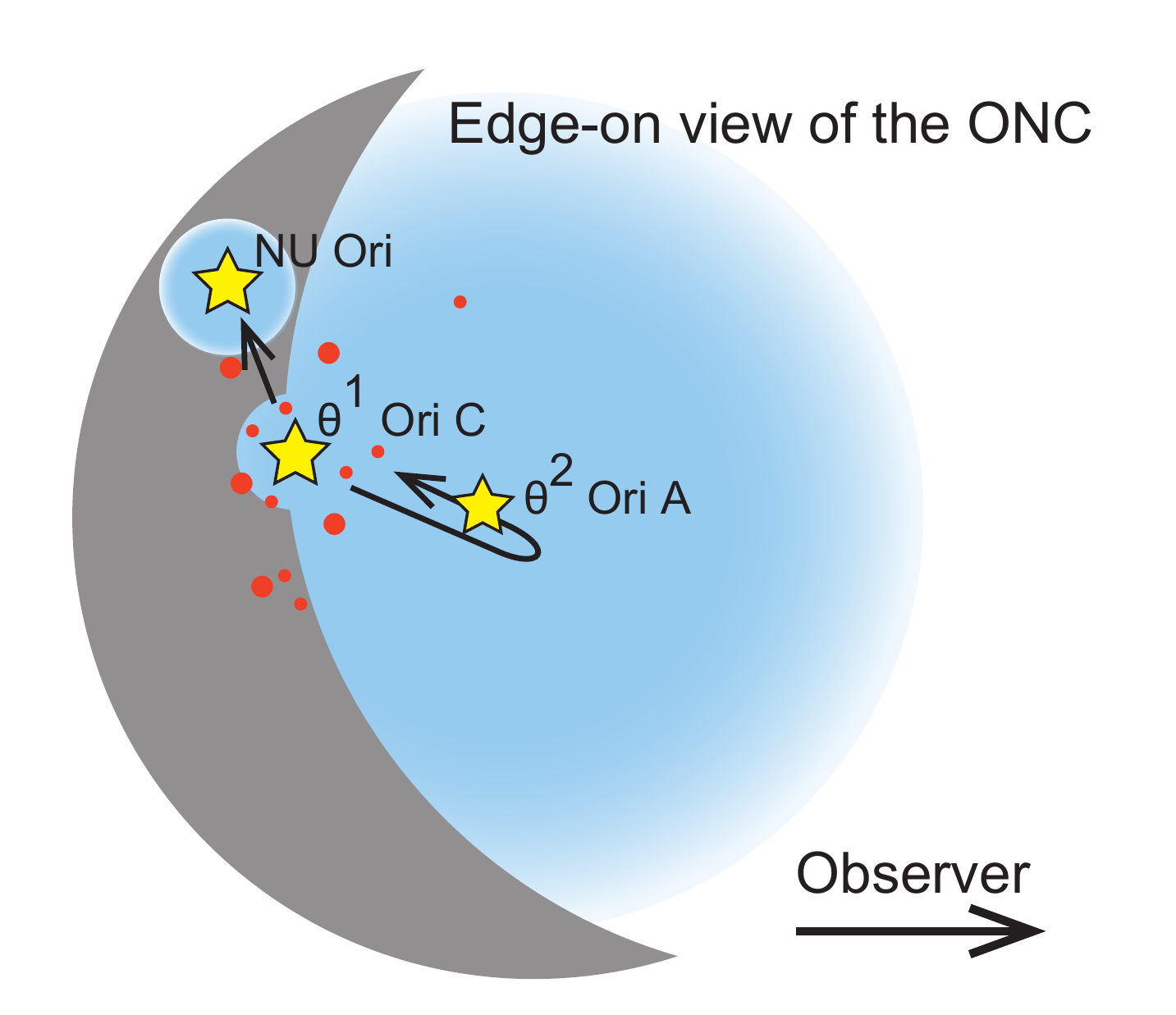}
 \end{center}
\caption{Schematic of the Orion Nebula structure. The gray region indicates a dense Orion molecular cloud. The star symbols indicate the O-type stars, and oranges indicates other stars. The blue spheres indicate the H{\sc ii} regions.}\label{fig:ONC_sche}
\end{figure}

The structure of the ONC is shown in Fig.~\ref{fig:ONC_sche}. $\theta^2$ Ori A was ejected from the ONC $<0.5$\,Myr ago and now is returning to the ONC center. Through the hole opened by the ejection of $\theta^2$ Ori A, $\theta^1$ Ori C can irradiate the gas toward the observer \citep{2008Sci...319..309G,2019Natur.565..618P}. Both $\theta^1$ Ori C and $\theta^2$ Ori A contribute to the formation of the 2-pc bubble as is suggested from the observations of this region \citep{2017ApJ...837..151O}.

We also calculate the bubble size and the trajectory of another O-star in the Orion Nebula,
NU Ori, which is located in the center of M43 (see Fig.~\ref{fig:fig1}). The mass of this star is estimated to be $16\pm3 M_{\odot}$ \citep{2018A&A...620A.116G}, and the bubble radius around it is $\sim 0.3$\,pc. The bubble is still embedded in the dense molecular cloud. The velocity relative to the ONC is 2.8\,km\,s$^{-1}$ in projection, and therefore we can estimate that this star has traveled from the ONC center. From the velocity and distance, we estimated that this star was ejected 0.1--0.2 Myr ago. The distance to NU Ori ($415^{+11}_{-10}$\,pc) suggests that this star appears to be behind the ONC, where the dense molecular gas still exists. Using equation (\ref{eq:shell}), we estimated that the size of 0.3\,pc at 0.1\,Myr is achieved for the wind luminosity of NU Ori ($L_{\rm w}\sim 10^{33}$\,erg\,s$^{-1}$) and the gas density of $10^4$\,cm$^{-3}$. These results are also consistent with observations.

\begin{table*}
\caption{Stellar parameters}
\begin{tabular}{lcccccc}
\hline\hline
Name & $T_{\rm eff}$ & $\log L/L_{\odot}$ & $M$ & $v_{\inf}$ & $\dot{M}$ &$L_{\rm w}$ \\
     & (K) &   & ($M_{\odot}$) & (km\,s$^{-1}$)& ($M_{\odot}$\,yr$^{-1}$) & (erg\,s$^{-1}$) \\
\hline
$\theta^1$ Ori C & 39000 & 5.31 & 45 & 2500 & $4.0\times 10^{-7}$ & $7.9\times 10^{35}$\\
$\theta^2$ Ori A & 35000 & 4.93 & 39 & 2000 & $5.9\times 10^{-8}$ & $7.4\times 10^{34}$\\
NU Ori & 31000 & 4.42 & 16 & 1000 & $1.0\times10^{-8}$ & $3.1\times 10^{33}$\\
\hline
\end{tabular}\label{tb:stars}

From the left, Column 1 is the name of the star, and Columns 2, 3, and 4 are the effective temperature, luminosity, and mass \citep{2006A&A...448..351S}.
Column 5 is the terminal velocity of the stellar wind \citep{1990ApJ...361..607P,2018A&A...620A..89N}.
Column 6 is the mass-loss rate obtained using Equation (12) of \citet{2000A&A...362..295V} with the effective temperature ($T_{\rm eff}$), luminosity ($L$), mass (M), the ratio of terminal velocity ($v_{\rm inf}$), and escape velocity ($v_{\rm esc}$). We assumed that $v_{\rm inf}/v_{\rm esc}=2.6$.
Column 7 is the mechanical luminosity of the wind, $L_{\rm w}=0.5\dot{M}v_{\rm inf}^2$.
\end{table*}

\section{Summary}
We performed $N$-body/SPH simulations of star cluster formation. Our simulation included the proper integration of stellar orbits, which revealed that star clusters eject massive stars from the cluster center.
Such ejected massive stars break the wall of the dense molecular cloud in the direction they escaped. The hole in the dense molecular cloud results in the formation of off-center H{\sc ii} regions such as the H{\sc ii} region associated with the Orion Nebula. In contrast, the molecular gas is too dense to be completely ionized, and therefore the star formation is still on-going.

We found that the distribution of velocities of OB stars around the ONC obtained from Gaia astrometric data is identical to the velocity distribution obtained from our simulations. This result implies that these OB stars around the ONC are formed in the ONC and scattered in the ONC center. 
Some scattered massive stars did not escape from the cluster and fell back to the cluster center. The astrometric data obtained from Gaia and our orbit analysis suggest that $\theta^2$ Ori A is such a star. 
 We also conclude that NU Ori was ejected from the ONC center 0.1--0.2\,Myr ago.

The dynamical ejection of stars can occur in any star cluster that contains multiple massive stars and is crucial for their formation because it changes the timing of the ionization of gas \citep{Kroupa2018,Wang2019,2021PASJ...73.1074F}. In addition to the formation of a star cluster, ejected massive stars may affect star formation on a larger scale. The observed high runaway fractions of massive stars (20\,\%) \citep{1961BAN....15..265B} can be explained by the dynamical ejection from star clusters \citep{2011Sci...334.1380F}. Runaway massive stars travel from their origin (dense molecular clouds) to low-density environments and efficiently ionize the interstellar gas. This may trigger new star formation \citep{2020MNRAS.494.3328A}. Some of these massive stars die in supernova explosions $\sim100$\,pc ($\simeq30$\,[km\,s$^{-1}] \times 3$\,[Myr]) away from their origin, which is a 1-pc-size star cluster. This strong feedback can affect the evolution of the interstellar medium in the 100-pc scale, as has also been suggested for the Orion complex \citep{2020ApJ...902..122K}, and even larger galactic scale star formation \citep{2020MNRAS.494.3328A}.

\section*{Acknowledgements}
The authors thank Steven Rieder for providing the \textsc{amuse} script for surface density and temperature map, Kurumi Ishikura and Ryoichi Nishi for discussion on the stellar distribution around the ONC, Takaaki Takeda (4D2U at the National Astronomical Observatory of Japan) for the visualization of the simulation, and Editage (www.editage.com) for English language editing. 
Numerical computations were carried out on Cray
XC50 CPU-cluster at the Center for Computational Astrophysics (CfCA) of the National Astronomical Observatory of Japan. 
This work was supported by JSPS KAKENHI Grant Number 19H01933, 20K14532, 21J00153, 21K03614, 21K03633, 21H04499 and Initiative on Promotion of Supercomputing for Young or Women Researchers, Information Technology Center, The University of Tokyo, and MEXT as “Program for Promoting Researches on the Supercomputer Fugaku” (Toward a unified view of the universe: from large scale structures to planets, Revealing the formation history of the universe with large-scale simulations and astronomical big data).
MF was supported by The University of Tokyo Excellent Young Researcher Program.
L.W. thanks the financial support from JSPS International Research Fellow (Graduate School of Science, The University of Tokyo). 
L.W. also thanks the support from the one-hundred-talent project of Sun Yat-sen University
and the National Natural Science Foundation of China through grant 12073090.

\section*{Data Availability}

\textsc{petar} is available here: {\tt https://github.com/lwang-astro/PeTar}.

\textsc{celib} is available here: {\tt https://bitbucket.org/tsaitoh/celib}.



\bibliographystyle{mnras}
\bibliography{example} 



\appendix
\section{OB stars near the ONC from Gaia catalog}

Table~\ref{tb:Gaia} lists the 34 OB stars within 5 pc (0.71$^{\circ}$) from the ONC.

\begin{table*}

\caption{Catalog of OB stars in the ONC region}
\begin{tabular}{llcccccc}
\hline
Name  & Sp. type  & R.A. & Dec. & Parallax & $(\mu_{\alpha*}, \mu_\delta)$ & $(\Delta v_{\alpha}, \Delta v_\delta)$ & $\Delta v$  \\
 &  & (deg) & (deg) &  (mas) & (mas yr$^{-1}$) & (km s$^{-1}$) & (km s$^{-1}$) \\ 
      \hline
Brun 508 & B9V & $+83.77$ & $-5.98$ & $2.62 \pm 0.03$ & $(+1.38, +0.53)$ & $(+0.51, +0.42)$ & $+0.65$ \\ 
HD  36919 & B9V & $+83.70$ & $-6.00$ & $2.64 \pm 0.03$ & $(+0.92, +0.07)$ & $(-0.33, -0.42)$ & $+0.53$ \\ 
* iot Ori B & B8III & $+83.86$ & $-5.91$ & $2.79 \pm 0.05$ & $(+1.13, +1.62)$ & $(+0.05, +2.25)$ & $+2.25$ \\ 
HD  37000 & B3/5 & $+83.80$ & $-5.93$ & $2.62 \pm 0.04$ & $(+1.44, -0.07)$ & $(+0.62, -0.67)$ & $+0.92$ \\ 
HD  36983 & B5(II/III) & $+83.78$ & $-5.87$ & $2.63 \pm 0.03$ & $(-0.35, +0.59)$ & $(-2.60, +0.53)$ & $+2.66$ \\ 
HD  36999 & B8(III) & $+83.81$ & $-5.83$ & $2.60 \pm 0.04$ & $(+1.34, +0.52)$ & $(+0.44, +0.39)$ & $+0.59$ \\ 
HD  36917 & B9III/IV & $+83.70$ & $-5.57$ & $2.22 \pm 0.06$ & $(+2.77, -1.74)$ & $(+3.57, -4.36)$ & $+5.63$ \\ 
HD  36939 & B7/8II & $+83.73$ & $-5.51$ & $2.38 \pm 0.04$ & $(+1.04, +0.64)$ & $(-0.12, +0.67)$ & $+0.68$ \\ 
HD  37150 & B3III/IV & $+84.06$ & $-5.65$ & $2.66 \pm 0.05$ & $(+1.21, -0.15)$ & $(+0.19, -0.80)$ & $+0.82$ \\ 
HD  37174 & B9V & $+84.11$ & $-5.41$ & $2.63 \pm 0.02$ & $(+1.27, +0.46)$ & $(+0.31, +0.29)$ & $+0.42$ \\ 
V* V1073 Ori & B9.5V & $+83.87$ & $-5.44$ & $2.61 \pm 0.03$ & $(-0.11, +1.01)$ & $(-2.19, +1.29)$ & $+2.54$ \\ 
HD  36982 & B1.5Vp & $+83.79$ & $-5.46$ & $2.45 \pm 0.02$ & $(+1.62, +1.78)$ & $(+1.00, +2.87)$ & $+3.04$ \\ 
* tet02 Ori C & B4V & $+83.88$ & $-5.42$ & $2.45 \pm 0.04$ & $(+2.51, +3.73)$ & $(+2.73, +6.63)$ & $+7.17$ \\ 
* tet02 Ori B & B2-B5 & $+83.86$ & $-5.42$ & $2.39 \pm 0.05$ & $(+1.16, +0.16)$ & $(+0.12, -0.27)$ & $+0.30$ \\ 
* tet02 Ori A & O9.5IVp & $+83.85$ & $-5.42$ & $2.97 \pm 0.21$ & $(+1.09, +2.39)$ & $(-0.02, +3.34)$ & $+3.34$ \\ 
* tet01 Ori D & B1.5Vp & $+83.82$ & $-5.39$ & $2.28 \pm 0.03$ & $(+1.82, +0.39)$ & $(+1.50, +0.19)$ & $+1.51$ \\ 
* tet01 Ori C & O7Vp & $+83.82$ & $-5.39$ & $2.50 \pm 0.14$ & $(+2.26, +0.99)$ & $(+2.20, +1.32)$ & $+2.56$ \\ 
* tet01 Ori A & B0V & $+83.82$ & $-5.39$ & $2.64 \pm 0.07$ & $(+1.36, +0.25)$ & $(+0.46, -0.09)$ & $+0.47$ \\ 
V* V1230 Ori & B1 & $+83.84$ & $-5.36$ & $2.46 \pm 0.03$ & $(+3.06, -1.42)$ & $(+3.79, -3.33)$ & $+5.05$ \\ 
NU Ori & O9V & $+83.88$ & $-5.27$ & $2.41 \pm 0.06$ & $(+0.92, +1.72)$ & $(-0.35, +2.79)$ & $+2.82$ \\ 
HD  36655 & B9V & $+83.28$ & $-5.34$ & $2.82 \pm 0.04$ & $(+0.11, +0.19)$ & $(-1.67, -0.18)$ & $+1.68$ \\ 
HD  36981 & B7III/IV & $+83.78$ & $-5.20$ & $2.59 \pm 0.04$ & $(+0.91, +0.37)$ & $(-0.34, +0.13)$ & $+0.36$ \\ 
HD  37060 & (B9) & $+83.89$ & $-5.11$ & $2.58 \pm 0.02$ & $(+1.40, +0.79)$ & $(+0.54, +0.90)$ & $+1.05$ \\ 
HD  37059 & B8/A0V & $+83.88$ & $-4.90$ & $2.60 \pm 0.03$ & $(+1.71, +0.81)$ & $(+1.12, +0.93)$ & $+1.45$ \\ 
HD  37058 & B3/5II & $+83.89$ & $-4.84$ & $2.62 \pm 0.04$ & $(+1.49, +0.74)$ & $(+0.71, +0.80)$ & $+1.07$ \\ 
HD 294264 & B3 & $+83.81$ & $-4.86$ & $2.43 \pm 0.05$ & $(+1.42, -1.44)$ & $(+0.63, -3.39)$ & $+3.45$ \\ 
* c Ori & B1V & $+83.85$ & $-4.84$ & $6.52 \pm 1.52$ & $(-4.12, -1.97)$ & $(-3.80, -1.65)$ & $+4.14$ \\ 
HD  36938 & B9V & $+83.73$ & $-4.77$ & $2.51 \pm 0.03$ & $(+1.99, -0.71)$ & $(+1.68, -1.90)$ & $+2.54$ \\ 
HD  36958 & B3/5V & $+83.77$ & $-4.73$ & $2.76 \pm 0.08$ & $(+0.97, +0.11)$ & $(-0.22, -0.32)$ & $+0.39$ \\ 
HD  37130 & B8/9IV & $+84.01$ & $-4.75$ & $2.56 \pm 0.02$ & $(+0.22, -1.01)$ & $(-1.63, -2.42)$ & $+2.92$ \\ 
\hline
HD  37025 & B3(III) & $+83.82$ & $-6.03$ & $2.65 \pm 0.08$ & $(+1.39, +2.04)$ & $(+0.52, +3.11)$ & $+3.15$ \\ 
HD  36960 & B1/2Ib/II & $+83.76$ & $-6.00$ & $2.62 \pm 0.12$ & $(+1.11, +1.68)$ & $(+0.02, +2.50)$ & $+2.50$ \\ 
HD  36959 & B1.2 & $+83.75$ & $-6.01$ & $2.79 \pm 0.11$ & $(+0.96, -1.00)$ & $(-0.24, -2.21)$ & $+2.23$ \\ 
HD  36918 & B8.3 & $+83.70$ & $-6.01$ & $2.66 \pm 0.03$ & $(+1.20, +0.51)$ & $(+0.19, +0.37)$ & $+0.41$ \\ 
      \hline
    \end{tabular}\label{tb:Gaia}
\\
From the left, columns 1 and 2 represent the star name and the spectral type in SIMBAD, respectively. 
Column 3 represents the right ascension.
Column 4 represents the declination.
Column 5 represents the parallax and its uncertainty in Gaia EDR3.
Column 6 represents the proper motion vector in Gaia EDR3.
Columns 7 and 8 represent the velocity vector with respect to the ONC's center-of-mass and its magnitude, respectively. 
HD 37025--HD 36918 are located within 5\,pc from the ONC in projection but out of the region shown in the panel (c) of Figure~\ref{fig:fig1}.
\end{table*}

In the following, we summarize the details of massive stars ($>10 M_{\odot}$) which are located in the Orion Nebula (M42 and M43). 
Here, we assumed the distance to the ONC and the radial velocity of the ONC as $388\pm5$\,pc \citep{2017ApJ...834..142K} and $27.45^{+0.21}_{-0.22}$\,km\,s$^{-1}$ \citep{2021arXiv210505871T}, respectively.

\subsection*{$\theta^1$ Ori C}
\begin{figure*}
 \begin{center}
 \includegraphics[width=120mm]{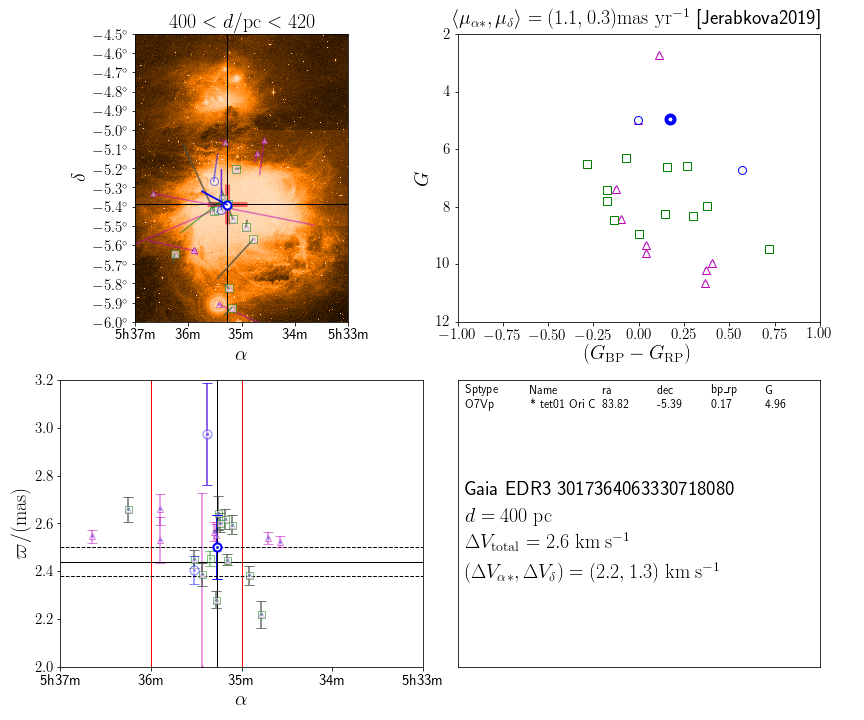}
 \end{center}
\caption{Distribution of bright stars ($G<12$ mag) near the ONC for which the stellar parallax data is consistent with $400 \;\mathrm{pc} < d < 420 \;\mathrm{pc}$ 
within 3 sigma level. 
The blue circles and green squares represent stars classified in SIMBAD 
as O- and B-type stars, respectively. 
The magenta triangles represent other bright stars with 
relatively blue color $(G_\mathrm{BP}-G_\mathrm{RP})<0.5$. 
The thick blue circle corresponds to $\theta^1$ Ori C. 
(Top left)
The relative proper motion vector of stars 
with respect to the ONC's systemic proper motion in \citet{2019A&A...627A..57J}. 
(Top right) 
The color-magnitude diagram. 
(Bottom left)
The parallax and its uncertainty. 
(Bottom right)
Summary of the observational property of $\theta^1$ Ori C and its tangential velocity in unit of $\rm{km\;s^{-1}}$ 
assuming a fiducial distance of $d = 400 \;\rm{pc}$.
}\label{fig:theta01_Ori_C}
\end{figure*}

$\theta^1$ Ori C (HD 37022) is the most massive star in the ONC. The parallax from Gaia EDR3 is $2.50\pm 0.14$, which is $400^{+24}_{-21}$\,pc (379--424 pc). The radial velocity is $24.50\pm 1.2$ km\,s$^{-1}$\citep{1972ApJ...174L..79C}. The proper motion relative to the ONC is 2.56 km\,s$^{-1}$. This value is similar to the velocity dispersion of stars in the ONC core measured with Gaia \citep{2021arXiv210505871T}. From these data, we consider that $\theta^1$ Ori C is inside the ONC. In Fig.~\ref{fig:theta01_Ori_C}, we visualized the data of this star.

We calculated the mass-loss rate using Equation (12) of \citet{2000A&A...362..295V} with the stellar parameters summarized in Table~\ref{tb:stars} and obtained $4.0\times 10^{-7} M_{\odot}$\,s$^{-1}$. The obtained mechanical luminosity was $7.9\times 10^{35}$\,erg\,s$^{-1}$, which is consistent with previous research (7--8$\times 10^{35}$\,erg\,s$^{-1}$) \citep{2008Sci...319..309G,2019Natur.565..618P}.

\subsection*{$\theta^2$ Ori A}
\begin{figure*}
 \begin{center}
 \includegraphics[width=120mm]{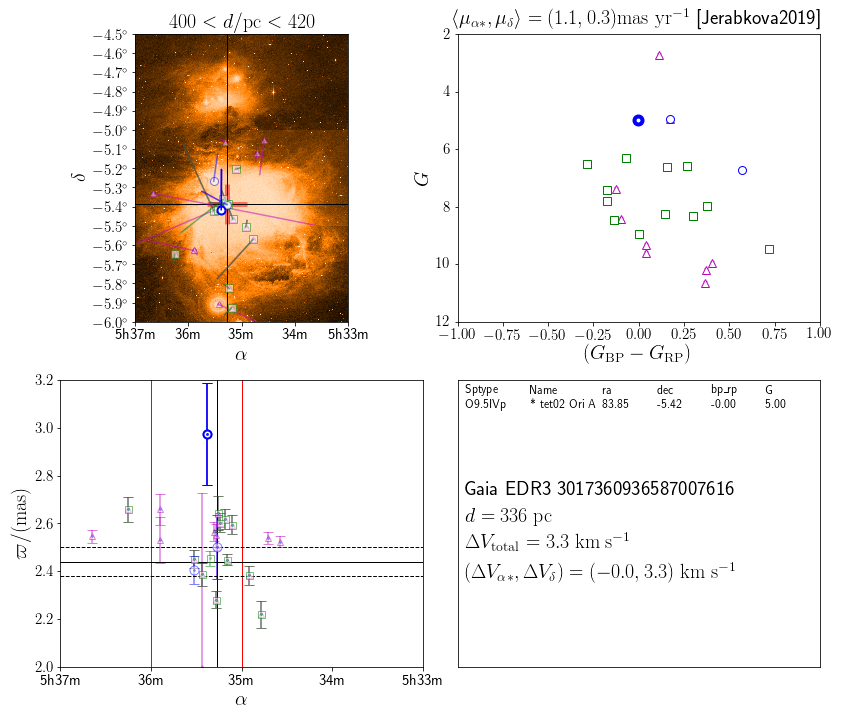}
 \end{center}
\caption{Same as Fig. \ref{fig:theta01_Ori_C} but for $\theta^2$ Ori A.}\label{fig:theta02_Ori_A}
\end{figure*}

$\theta^2$ Ori A (HD 37041) is the second brightest star in this region. The estimated mass is 25--39\,$M_{\odot}$ \citep{1999NewA....4..531P,2006A&A...448..351S}.
The parallax is $2.97\pm 0.21$, which corresponds to the distance of $336\pm^{+26}_{-22}$\,pc (314--362). This star is located in front of the ONC, albeit the distance is relatively uncertain (see Fig.~\ref{fig:theta02_Ori_A}).
The radial velocity is 20--30\,km\,s$^{-1}$  \citep{1920PDO.....5....1H,1974JRASC..68..205A}.

From the distance including the error, this star is at least 20\,pc closer to observer with $1\sigma$ error. However, the parallax error of this star is $\sim 10$\,\%, and the error of 0.1 mas in parallax corresponds to 16\,pc at a distance of 400\,pc. Therefore, the distance of $\theta^2$ Ori A to the ONC can be shorter.
The projected distance of $\theta^2$ Ori A to the ONC is $\sim 0.3$\,pc. The projected relative velocity is 3.34 km\,s$^{-1}$, and the direction of the velocity is toward the ONC center. 

We also estimated the mechanical luminosity of $\theta^2$ Ori A. The mass-loss rate is obtained to be $5.9\times 10^{-8} M_{\odot}$\,yr$^{-1}$ with the stellar parameters summarized in Table~\ref{tb:stars}. We assume the wind terminal velocity as 2000\,km\,s$^{-1}$ \citep{1990ApJ...361..607P}. The estimated mechanical luminosity is $7.4\times 10^{34}$\,erg\,s$^{-1}$, which is an order of magnitude smaller than that of $\theta^1$ Ori C.

\subsection*{NU Ori}
\begin{figure*}
 \begin{center}
 \includegraphics[width=100mm]{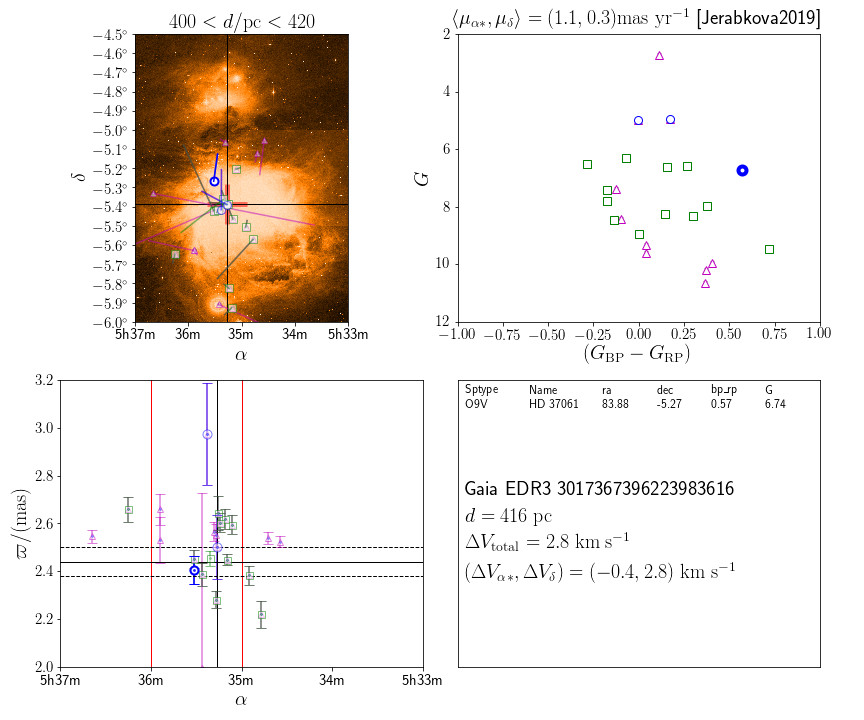}
 \end{center}
\caption{Same as Fig. \ref{fig:theta01_Ori_C} but for NU Ori (HD 37061).}\label{fig:HD37061}
\end{figure*}

NU Ori (HD 37061) is an O9V or B0.5V type star at the center of the small HII region in M43 \citep{2011A&A...530A..57S}. The parallax is $2.41\pm0.06$. The corresponding distance is $415^{+11}_{-10}$ (405--426\,pc).
The observed radial velocity is 66.90 km/s (SIMBAD) \citep{1970ApJS...19..387A}, which suggests $>30$\,km\,s$^{-1}$ to the ONC. The proper motion relative to the ONC is 2.82 km\,s$^{-1}$, and it is apparently escaping from the ONC (see Figs.~\ref{fig:fig1} and \ref{fig:HD37061}). This suggests that this star may be a runaway star returning toward the ONC. 

The mass is estimated to be 13--16$M_{\odot}$ \citep{2018A&A...620A.116G}. From the stellar parameters given in Table~\ref{tb:stars}, we obtained $L_{\rm w}=3.1\times10^{33}$\,erg\,s$^{-1}$.

\section{Simulation without gravitational softening}
We also confirmed that the dynamical ejections of massive stars are crucial for the formation of star clusters by performing a reference run with a gravitational softening, with which the strong dynamical encounters among stars are `softened' and as a result the ejection of stars are suppressed. 

\begin{figure*}
 \begin{center}
 \includegraphics[width=80mm]{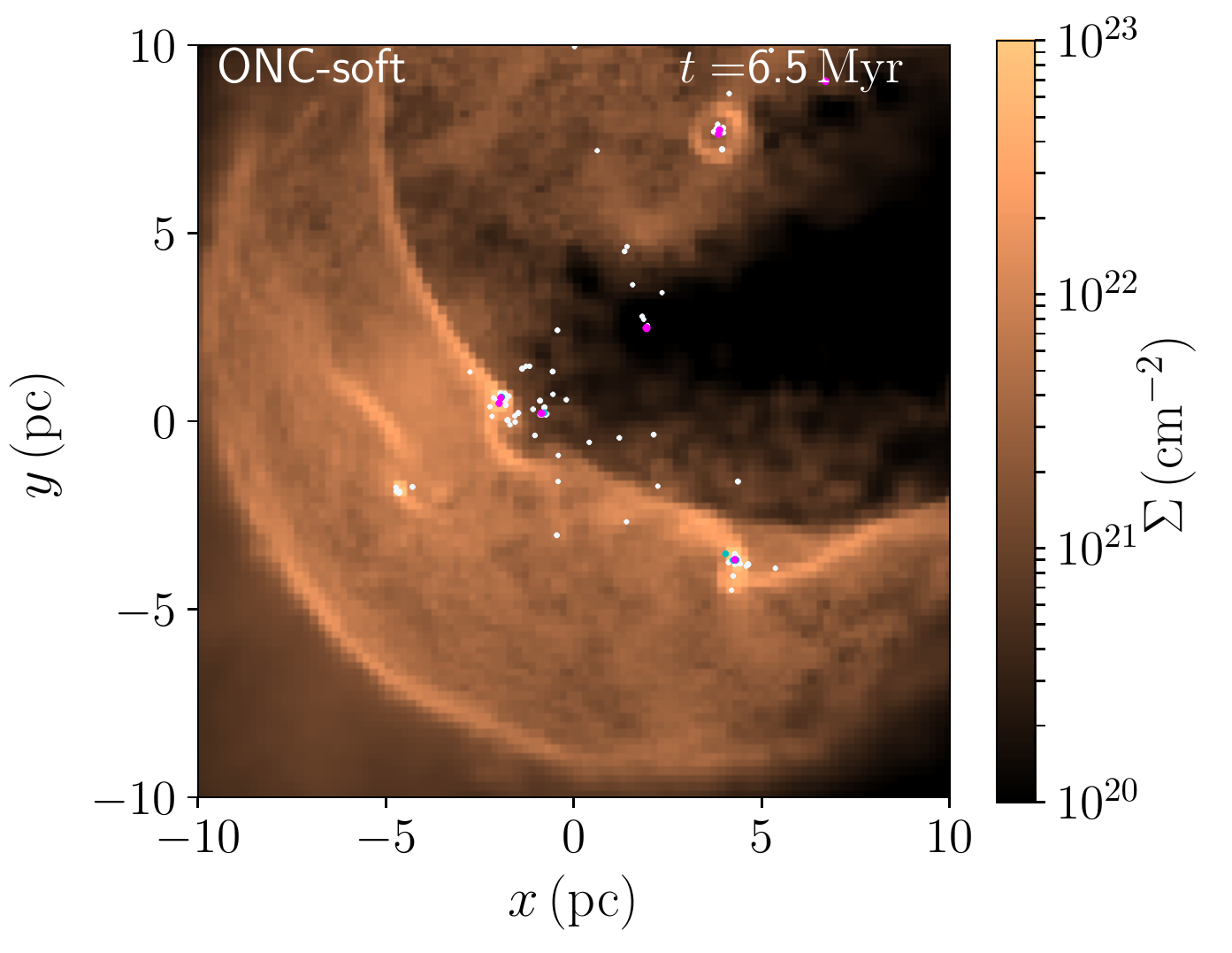}
 \includegraphics[width=80mm]{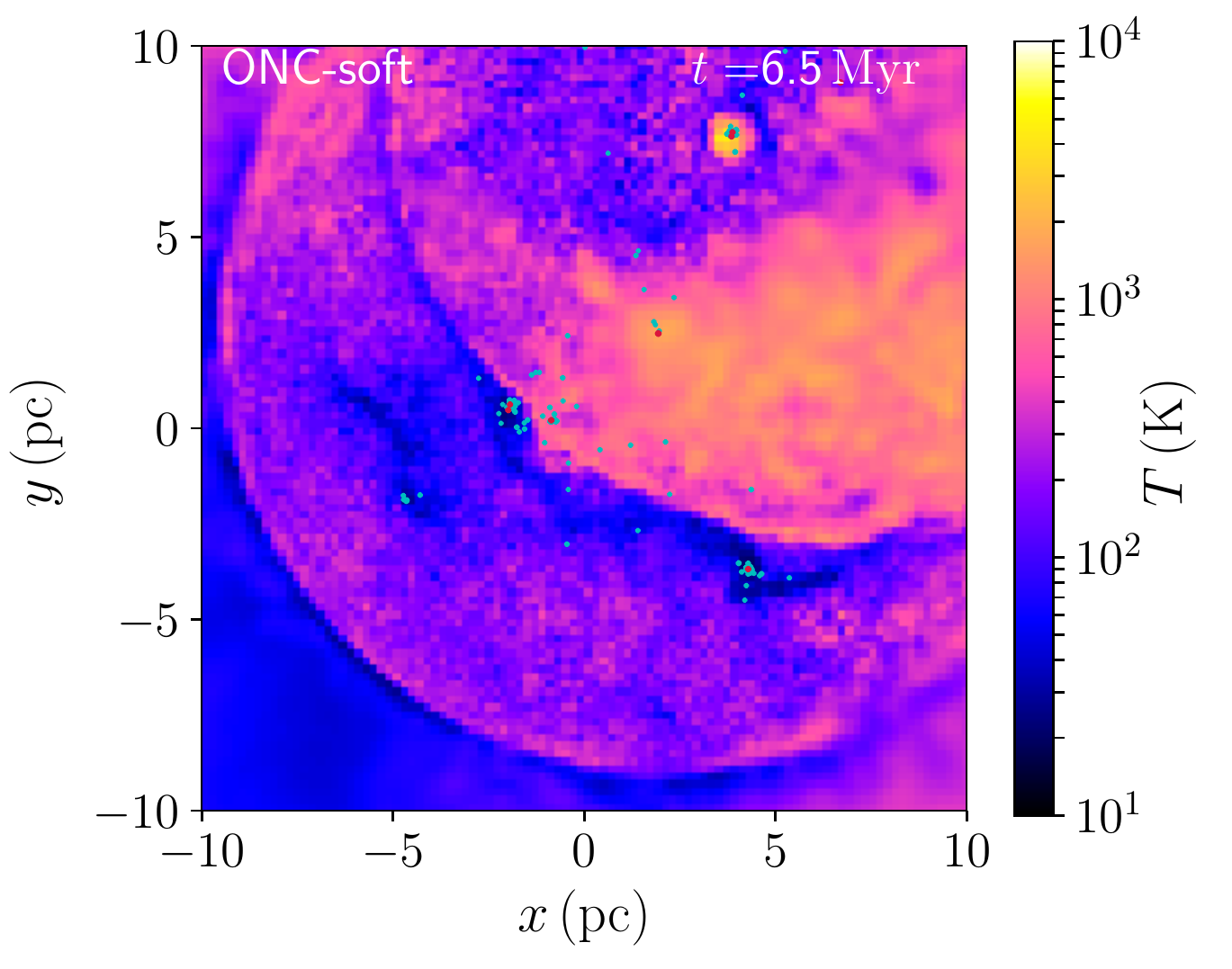}
 \end{center}
\caption{Snapshots of the simulation with a softening. The softening length is 0.07\,pc, which is the same as that for gas particles. Time indicates the time from the beginning of the simulation.}\label{fig:snapshot_soft}
\end{figure*}

We additionally performed a simulation with gravitational softening for stars. In this simulation, the initial condition of gas is exactly the same  as Model A. The softening length for stars was set to be the same as that for gas particles (0.07\,pc). The snapshots at $t_{\rm sim}=$ 6.5\,Myr is shown in Fig.~\ref{fig:snapshot_soft}. Unlike in the case without softening (see Fig.~\ref{fig:snapshot}), some small stellar clumps begin ionizing the surrounding gas from inside the clumps. This prevents them from evolving into a massive cluster via mergers of the clumps. With a gravitational softening, massive stars born in small clumps are kept inside the clumps without being ejected to the outside of the clumps. Therefore, the feedback inside the clumps works more efficiently when compared with the case without softening. We observed the suppression of stellar feedback owing to the ejection of massive stars. This was discussed in \citet{Kroupa2018}.

\bsp	
\label{lastpage}
\end{document}